\documentstyle[aps,preprint,amssymbols]{revtex}
\tighten
\begin{document}
\draft
\preprint{UTPT-95-09, gr-qc-9505005}

\title{Massive NGT and Spherically Symmetric Systems}
\author{M. A. Clayton}
\address{Department of Physics, University of Toronto, Toronto,
ON., Canada, M5S
1A7}
\date{\today}


\maketitle

\begin{abstract}
The arguments leading to the introduction of the massive
Nonsymmetric
Gravitational action are reviewed
\cite{Moffat:1994,Moffat:1995b},
leading to an action that gives asymptotically well-behaved
perturbations on GR
backgrounds.
Through the analysis of spherically symmetric perturbations about
GR
(Schwarzschild) and NGT (Wyman-type) static backgrounds, it is
shown that
spherically symmetric systems are not guaranteed to be static,
and hence
Birkhoff's theorem is not valid in NGT.
This implies that in general one must consider time dependent
exteriors when
looking at spherically symmetric systems in NGT.
For the surviving monopole mode considered here there is no
energy flux as it is
short ranged by construction.
Further work on the spherically symmetric case will be motivated
through a
discussion of the possibility that there remain additional modes
that do not
show up in weak field situations, but nonetheless exist in the
full theory and
may again result in bad global asymptotics.
A presentation of the action and field equations in a general
frame is given in
the course of the paper, providing an alternative approach to
dealing with the
algebraic complications inherent in NGT, as well as offering a
more general
framework for discussing the physics of the antisymmetric sector.
\end{abstract}


\section{Introduction}
\label{section:introduction}
In General Relativity (GR), just as in Maxwell's electrodynamics,
one finds that
given a spherically symmetric system, there are no dynamical
degrees of freedom
in the theory.
This is Birkhoff's theorem, and implies that a time dependent
source will not
excite modes in the gravitational system, so that outside this
source the system
must be physically equivalent to the Schwarzschild solution (the
non-trivial
static spherically symmetric spacetime).
It has been further established
\cite{Regge+Wheeler:1957,Vishveshwara:1970} that
the solution is stable when perturbed, so that small deviations
from spherical
symmetry do not alter the large scale features of the spacetime,
and systems
that are only approximately spherically symmetric are therefore
still very well
modeled by the Schwarzschild solution.
This establishes that the phenomenology of the Schwarzschild
solution is
physically relevant.

To see that Birkhoff's theorem is not a generic feature of
physical theories one
need look no further than a scalar field.
However the example that is important for this work is the
(massive) Kalb-Ramond
action \cite{Kalb+Ramond:1974} which (as will be demonstrated in
Section
\ref{section:Schw pert}) has a single mode that is time dependent
in general.
As will be shown in Section \ref{section:Massive NGT}, the
massive Nonsymmetric
Gravitational Theory \cite{Moffat:1994,Moffat:1995b} (mNGT as
opposed to the
older versions of the theory, referred to as NGT, or massless
NGT) becomes
identically a massive Kalb-Ramond field with an additional
curvature coupling
term when considered as a perturbation about a Ricci-flat GR
background, so the
result that NGT has a monopole mode is not surprising.
The mode considered here is short ranged, so that far enough away
from the
source one finds that the solution will be dominated by
Schwarzschild behavior,
and there is no energy flux.
However, after demonstrating that spherically symmetric fields in
the skew
sector are not static in general, it will be shown in Section
\ref{section:Wyman
pert} that the symmetric sector will also no longer be static,
through
examination of a similar perturbation about the mNGT background
discussed in
Section \ref{section:Wyman}.
This means that no static solutions can be considered rigorously
as an exterior
solution unless the solution is globally static (i.e. the
interior is static as
well).

The results in this paper are obtained through an examination of
linearized
perturbations, although the conclusions must hold in general as
Birkhoff's
theorem would imply that these fields must be static as well.
What cannot be examined in this fashion is whether in the full
nonlinear theory
more modes become excited.
In particular, one will see in the case of a perturbation about a
GR background
(in Section \ref{section:Massive NGT}) that there are three
propagating modes,
even though the absence of gauge invariant kinetic terms in the
full theory
would suggest that all six modes in the antisymmetric metric
could be
independent degrees of freedom (as yet, the number of degrees of
freedom in mNGT
has not been rigorously established).
Although this issue is not addressed directly in this work, the
ability to
recast the theory in a general basis given in Section
\ref{section:GenBasis
intro} sets the stage for a complete analysis of the spherically
symmetric
system in NGT.
Given that there is an additional mode in the general spherically
symmetric
system, how the fields may or may not approach an asymptotically
flat spacetime
should then be addressed, and also whether evolution
singularities of the type
discussed in \cite{Isenberg+Nestor:1977} are encountered.

\section{Massive NGT}
\label{section:Massive NGT}

The original version of NGT \cite{Moffat:1979} grew out of a
re-interpretation
of the Einstein-Straus \cite{Einstein:1945,Einstein+Straus:1946}
unified field
theory as a purely gravitational system.
The antisymmetric part of the metric and connection operationally
produce
different modes of parallel transport and index contraction
\cite{Moffat:1979,Moffat:1990,Einstein:1956}, where the algebra
is consistent
with an enlargement of the tangent vector space to its hyperbolic
complex
extension
\cite{Kunstatter+Yates:1981,Mann:1989,Kunstatter+Moffat+Malzan:1983}.
It is important to note that the action cannot support the
additional Bianchi
identities and gauge invariance related to the extension of the
tangent bundle,
simply because the base manifold is locally diffeomorphic to
${\cal R}^4$,
and the variational principle is based on an integration over
this real
manifold.
Any change of gauge that mixes real and hyperbolic complex
covectors will cause
the volume element to pick up a hyperbolic complex piece, and the
action will no
longer be real.

The hyperbolic complex structure is unnecessary for an
operational discussion of
the theory (although it may be relevant for a more fundamental
discussion of its
physical interpretation), and in this paper all quantities will
be considered
real, allowing antisymmetric contributions to the metric and
connection
coefficients.
The dynamics of the theory will be determined from the first
order action
($G=c=1$):
\begin{equation}\label{mNGTAct}
S=\int d^4x\{-{\bf
g}^{\mu\nu}R_{\mu\nu}^{\text{\tiny{NS}}}[\Gamma]
-{\bf g}^{\mu\nu}\partial_{[\mu}W_{\nu]}
+{\bf l}^\mu\Gamma_\mu
+\case{1}{2}\alpha{\bf g}^{(\mu\nu)}W_\mu W_\nu
+\case{1}{4}m^2{\bf g}^{[\mu\nu]}g_{[\mu\nu]}\}+S_M,
\end{equation}
where:
\begin{equation}\label{mMAct}
\frac{\delta S_M}{\delta g^{\mu\nu}}={\bf T}_{\mu\nu},
\end{equation}
is the matter stress energy tensor that acts as a source in the
gravitational
field equations.
The Ricci-like tensor for NGT appearing in (\ref{mNGTAct}) is
written as:
\begin{equation}\label{NGTRicci}
R_{\mu\nu}^{\text{\tiny{NS}}}=
\partial_\alpha\Gamma^\alpha_{\mu\nu}-\case{1}{2}
(\partial_\nu\Gamma^\alpha_{\mu\alpha}+\partial_\mu\Gamma^\alpha_
{\alpha\nu})
+\Gamma^\alpha_{\mu\nu}\Gamma^\beta_{(\alpha\beta)}
-\Gamma^\alpha_{\beta\nu}\Gamma^\beta_{\mu\alpha},
\end{equation}
and a mass term for $g_{[\;\;]}$ has been included along with a
term quadratic
in $W$ as new features of the action.
As will become clear shortly, $\alpha$ may be fixed uniquely by
requiring good
asymptotic behavior of perturbations about GR backgrounds.
The new parameter in the massive action ($m$) is an inverse
length scale that
must be constrained by experiment, and $l$ is a Lagrange
multiplier employed to
enforce the vanishing of the trace of the antisymmetric part of
the connection
coefficients: $\Gamma_\mu=\Gamma^\alpha_{[\mu\alpha]}$.

The field equations related to the metric compatibility
conditions are derived
through the variations:
\begin{mathletters}
\label{mVars}
\begin{eqnarray}
\label{mVars:a}
\frac{\delta S}{\delta {\bf l}^\mu}&=&\Gamma_\mu=0,  \\
\label{mVars:b}
\frac{\delta S}{\delta W_\mu}&=&\partial_\nu{\bf
g}^{[\nu\mu]}+\alpha{\bf
g}^{(\mu\nu)}W_\nu=0,  \\
\label{mVars:c}
\frac{\delta S}{\delta \Gamma^\gamma_{\sigma\omega}}&=&
\partial_\gamma{\bf g}^{\sigma\omega}
-{\bf g}^{\sigma\omega}\Gamma^\alpha_{(\gamma\alpha)}
+{\bf g}^{\alpha\omega}\Gamma^\sigma_{\alpha\gamma}
+{\bf g}^{\sigma\alpha}\Gamma^\omega_{\gamma\alpha}\nonumber  \\
&&\quad -\case{1}{2}\delta^\omega_\gamma
(\partial_\alpha{\bf g}^{\sigma\alpha}
+{\bf g}^{\alpha\beta}\Gamma^\sigma_{\alpha\beta}
-{\bf l}^\sigma)
-\case{1}{2}\delta^\sigma_\gamma
(\partial_\alpha{\bf g}^{\alpha\omega}
+{\bf g}^{\alpha\beta}\Gamma^\omega_{\alpha\beta}
+{\bf l}^\omega).
\end{eqnarray}
\end{mathletters}
Contracting on either index of (\ref{mVars:c}) and solving for
the
(anti-)symmetric parts of the divergence of the densitized
inverse metric
results in the determination of the Lagrange multiplier using
(\ref{mVars:b}):
${\bf l}^\sigma=\frac{\alpha}{3}{\bf
g}^{(\sigma\omega)}W_\omega$.
This also allows one to simplify the Kronecker-$\delta$ terms,
and determine the
compatibility conditions in undensitized form as:
\begin{equation}\label{Comp2m}
\partial_\gamma g_{\mu\nu}
-g_{\mu\alpha}\Gamma^\alpha_{\nu\gamma}
-g_{\alpha\nu}\Gamma^\alpha_{\gamma\mu}
=\case{2}{3}\alpha
(g_{\mu[\gamma}g_{\alpha]\nu}
+\case{1}{2}g_{\mu\nu}g_{[\gamma\alpha]})g^{(\alpha
\beta)}W_\beta,
\end{equation}
where the inverse of the metric has been defined by
$g_{\mu\alpha}g^{\alpha\nu}=\delta^\nu_\mu$, which has been used
in order to
rewrite the compatibility conditions in terms of the components
of the metric
$g_{\mu\nu}$.

The remaining field equations derived from the variation of the
action with
respect to $g^{\mu\nu}$ may be written as:
\begin{equation}\label{mNGTFeq}
{\cal R}_{\mu\nu}:=R_{\mu\nu}^{\text{\tiny{NS}}}
+\partial_{[\mu}W_{\nu]}-\case{1}{2}\alpha W_{\mu}W_{\nu}
-\case{1}{4}m^2(g_{[\mu\nu]}
-g_{\alpha\mu}g_{\nu\beta}g^{[\alpha\beta]}
+\case{1}{2}g_{\nu\mu}g^{[\alpha\beta]}g_{[\alpha\beta]})
=T_{\mu\nu}-\case{1}{2}g_{\nu\mu}T,
\end{equation}
where $T=g^{\mu\nu}T_{\mu\nu}$, and the tensor ${\cal R}$ has
been
introduced in order to simplify the discussion of the field
equations.
One may translate the conventions used here to those in
\cite{Moffat:1994} by
taking $W\rightarrow-\frac{2}{3}W, T_{\mu\nu}\rightarrow -8\pi
T_{\mu\nu}$,
$\alpha\rightarrow -\frac{9}{4}\sigma$, and adjusting the
definitions of the
inverse metric: $g^{\mu\nu}\rightarrow g^{\nu\mu}$.
To see the equivalence of the action, one further needs to
rewrite $\Gamma$ in
terms of the unconstrained $W$ connection, and drop the
contribution from the
Lagrange multiplier $l$.

The action for massless NGT is given by (\ref{mNGTAct}) with
$m=\alpha=0$.
As will be demonstrated, the new terms have been introduced in
order to make all
skew modes short ranged when considering perturbations about GR
backgrounds.
One performs this expansion about a symmetric, Ricci-flat
background, where one
assumes that all background curvatures fall off at worst as 1/r
as
$r\rightarrow\infty$.
This allows one to talk sensibly of energy-momentum and decompose
fields via a
spin projection, so that higher order poles and negative energy
(ghost) modes
may be identified, as well as avoiding the full nonsymmetric
structure of a more
general background that would make the analysis far more
complicated.
One considers a perturbation of all quantities about a symmetric
GR background
as in \cite{Kelly:1991}:
\begin{eqnarray}\label{foexp}
g_{\mu\nu}&&\rightarrow g_{(\mu\nu)}+h_{\mu\nu}, \nonumber \\
\Gamma^\alpha_{\mu\nu}&&\rightarrow \{\stackrel{{\scriptstyle
\alpha}}{{\scriptstyle \mu\nu}}\}
+\gamma^\alpha_{\mu\nu},
\end{eqnarray}
where $W$, $l$ and $T$ are are considered to be first order in
the perturbation
(as the background is assumed to be Ricci-flat, there one has
$T_{\mu\nu}=0$).
As usual, indices will be `raised' and `lowered' by the symmetric
background
metric, and the covariant derivative $\nabla$ is associated with
the background
Christoffel symbols $\{\;\}$ determined from the background
metric in the usual
manner.
Corrections to the background curvatures, field equations and
Lagrangian at each
order in the perturbation will be indicated by a superscript as:
${}^0\! R, {}^1\! R\cdots$.

The first order correction to the compatibility equation
(\ref{Comp2m}) can be
solved explicitly for $\gamma$ to yield:
\begin{equation}\label{lcomp}
\gamma^\alpha_{\mu\nu}=\case{1}{2}g^{\alpha\beta}
(\nabla_\nu h_{\beta\mu}+\nabla_\mu h_{\nu\beta}-\nabla_\beta
h_{\mu\nu})
+\case{2}{3}\alpha\delta^\alpha_{[\mu}W_{\nu]},
\end{equation}
and $\gamma_\mu=\gamma^\alpha_{[\mu\alpha]}$ is seen to vanish by
the
linearization of the skew divergence equation:
\begin{equation}\label{ldiv}
-\nabla_\nu h^{[\mu\nu]}=\alpha W^\mu.
\end{equation}

In massless NGT \cite{Moffat:1990}, one had
(\ref{lcomp},\ref{ldiv}) with
$\alpha=0$, and hence there was no relation between the metric
degrees of
freedom and those of $W$.
The skew part of the linearization of equation (\ref{mNGTFeq})
with $m=0$ as
well as $\alpha=0$ became:
\begin{equation}\label{NGTlFeq}
{}^1\! {\cal R}_{[\mu\nu]}={}^1\!
R_{[\mu\nu]}+\partial_{[\mu}W_{\nu]}
=\nabla_\alpha\gamma^\alpha_{[\mu\nu]}+\partial_{[\mu}W_{\nu]}
=T_{[\mu\nu]}.
\end{equation}
The symmetric contribution, are the equations for a metric
perturbation in GR
\cite{Wald:1984}, and will be ignored in the remainder of this
section.
Using (\ref{ldiv}) with $\alpha=0$, the assumption that the
background is
Ricci-flat (${}^0\! R_{\mu\nu}=0$), and the commutation
relation (for
an arbitrary tensor $B$):
\begin{equation}\label{commcovder}
\nabla_{[\alpha}\nabla_{\beta]}B_{\mu\nu}
=-\case{1}{2}(B_{\omega\nu}{}^0\! R^\omega_{\;\;\mu\alpha\beta}
+B_{\mu\omega}{}^0\! R^\omega_{\;\;\nu\alpha\beta}),
\end{equation}
(\ref{NGTlFeq}) simplifies to:
\begin{equation}\label{lNGTf}
\nabla^\alpha\nabla_\alpha h_{[\mu\nu]}
-2\partial_{[\mu}W_{\nu]}
-4{}^0\! R^{\alpha\;\;\beta}_{\;\;\mu\;\;\nu}h_{[\alpha\beta]}
=\nabla^\alpha F_{\mu\nu\alpha}
-2\partial_{[\mu}W_{\nu]}-8{}^0\!
R^{\alpha\;\;\beta}_{\;\;\mu\;\;\nu}h_{[\alpha\beta]}
=-2T_{[\mu\nu]}.
\end{equation}
The second form is given in terms of the curl of the skew metric
($F_{\gamma\mu\nu}=\partial_\gamma h_{\mu\nu}+\partial_\nu
h_{\gamma\mu}+\partial_\mu h_{\nu\gamma}$), in order to more
easily demonstrate
the result found previously by Damour, Deser, and McCarthy
\cite{Damour+Deser+McCarthy:1992,Damour+Deser+McCarthy:1993}
(using the fact
that $\nabla^\gamma\nabla^\nu F_{\mu\nu\gamma}=0$ about a
Ricci-flat
background), that these antisymmetric perturbations will in
general have bad
asymptotic behavior.
Although written as if the skew metric were a gauge field, the
presence of the
curvature coupling term implies that the associated gauge
invariance is not
present \cite{Kelly:1992}, and one is not allowed to make any
choice of gauge in
order to simplify this sector.
One proceeds by taking the divergence of (\ref{lNGTf}) and
choosing the gauge
$\nabla^\alpha W_\alpha=0$ (the theory had a $U(1)$ invariance as
$W$ only
appeared in the action in a curl) to find:
\begin{equation}\label{lNGTdiv}
\nabla^\alpha\nabla_\alpha W_\mu
-8\nabla^\nu[{}^0\!
R^{\alpha\;\;\beta}_{\;\;\mu\;\;\nu}h_{[\alpha\beta]}]
=-2\nabla^\nu T_{[\mu\nu]}.
\end{equation}
Notice that the background curvature acts as a source here, so
that even if one
postulates that the matter source is conserved, this curvature
coupling (and in
general other nonlinear terms) will still exist as a source,
causing $W$ to
propagate and have asymptotic behavior consistent with a massless
field ($\sim
1/r$ along the forward light cone).
Using this asymptotic behavior to determine $h$ from
(\ref{lNGTf}) results in a
source with $\sim 1/r$ behavior, causing the field $h$ not to
fall off as
$r\rightarrow\infty$ along the forward light cone.
This analysis is correct since one has assumed that the
background curvature
falls off fast enough, and hence the potential term in
(\ref{lNGTdiv}) can be
treated as a source, without changing the asymptotic behavior of
the fields.
It must be stressed that one is assuming that the background {\em
and} the
radiative fields fall off at least as fast as $\sim 1/r$, and
what has actually
been derived here is a contradiction of this, since $h$ is driven
to a constant
and can no longer be considered as a perturbative mode.

Any analysis of this sort also supposes that a solution of the
linearized field
equations does in fact correspond to an exact solution of the
full nonlinear
field equations.
This is the case in GR \cite{Wald:1984,Fischer+Marsden:1979} at
least for
source free equations, but no such result exists yet for any of
the models
considered here.
It is possible to take the stance that NGT is not
linearization-stable, so that
this sort of analysis necessarily produces spurious results that
do not
correspond to global solutions, but then one is denying the
ability to do any
sort of perturbative analysis without the existence of an exact
solution to back
it up.
Due to the scarcity of solutions, and the apparent existence of
weak-field
perturbative situations, this would seem an unreasonable position
to adopt.

This result is not confined to curved backgrounds, and in fact
the analysis
about Minkowski space will serve to explicitly demonstrate the
higher order pole
leading to bad fall-off.
Since this curvature coupling, and any nonlinear effects in
general, will act as
a nonconserved source term in the skew sector, the linearization
that correctly
represents the full nonlinear field equations in the asymptotic
region will have
a nonconserved source term.
This is no more than the observation that once again, the full
NGT action does
not possess any form of additional gauge invariance in the skew
sector.
A gauge field coupled to a source (or matter) in a non-gauge
invariant manner
may have drastically different behavior than the empty space and
apparently
gauge invariant field equations, if indeed the action is
consistent at all.
Any analysis that attempts to determine the propagator or
asymptotic behavior of
the field must take the form of the source (or coupling to other
fields) into
account.

A trivial example of this is given by considering the Maxwell
action.
Coupling the usual gauge invariant kinetic terms to a
nonconserved source gives
an inconsistent set of field equations,
and adding some sort of gauge fixing term will give a consistent
set of
equations, but the scalar ghost mode will be excited and
depending on the gauge
there may be higher order poles in the solution.
The linearized field equations considered outside the source
resemble those of
the gauge invariant theory in a particular gauge, but treating
them as such will
not give asymptotic behavior that follows from coupling to the
nonconserved
source.
The situation in NGT is more akin to enforcing the gauge
condition in the action
through the use of an auxiliary field as: $b\partial_\mu A^\mu$
\cite{Nakanishi:1967,Goto+Obara:1967}.
Source conservation and absence of ghosts relies on whether or
not the scalar
Lagrange multiplier field $b$ has a source or not in the wave
equation that
determines it, and is thus a global question.
Given that the source for $A$ is not conserved, then $b$
propagates and there
are higher order poles in the solution for $A$, leading to fields
that do not
fall off as $r\rightarrow\infty$ along the forward light cone.

Considering the field equations for massless NGT linearized from
(\ref{lNGTf})
about Minkowski space:
\begin{mathletters}
\label{lNGTflat}
\begin{eqnarray}
\label{lNGTflat:a}
\square h_{[\mu\nu]}-2\partial_{[\mu}W_{\nu]}
&=&-2T_{[\mu\nu]}, \\
\label{lNGTflat:b}
\partial_\nu h^{[\mu\nu]}&=&0,
\end{eqnarray}
\end{mathletters}
($\partial^\nu T_{[\nu\mu]}\neq 0$),
one may take a divergence (to find a wave equation for $W$) or a
curl (to remove
$W$) of the first of these, resulting in the unique consistent
solution:
\begin{eqnarray}\label{LCh}
h_{[\mu\nu]}&=&-2\square^{-1}\left[T_{[\mu\nu]}
+2\square^{-1}\partial^\alpha\partial_{[\mu}T_{[\alpha\nu]]}
\right], \nonumber \\
W_\mu&=&-2\square^{-1}\partial^\nu T_{[\mu\nu]}.
\end{eqnarray}
The presence of the higher order pole (and consequent bad
fall-off) is now
obvious from the presence of the $\square^{-2}$ term in the Green
function
solution, and is no more than the result of vanNieuwenhuizen
\cite{vanNieuw:1973} who showed that the only healthy quadratic
actions built of
antisymmetric tensor fields are the so-called Kalb-Ramond
\cite{Kalb+Ramond:1974} (massless or massive) actions.
One also notes that there are 5 modes here: 3 in $h$, since 3 are
determined
algebraically by the second equation in (\ref{lNGTflat}), and 2
in $W$ due to
the previously mentioned $U(1)$ gauge invariance
\cite{Kunstatter+Leivo+Savaria:1984}.
If it is assumed that the (matter) source is conserved, the
higher order poles
are removed at linear order, but show up in the second order
correction to the
fields, again causing a breakdown of the perturbative analysis.

This analysis correctly represents the asymptotic behavior of the
fields
($W,h$), and is equivalent to equation (18) of
\cite{Mann+Moffat:1982}, where
the higher order pole resides in the projection operator:
$P(1^+)$.
One also sees the true propagating nature of $W$, and this is
borne out by the
analysis in \cite{Moffat:1980,McDow+Moffat:1982} where there are
five degrees of
freedom evolving from each Cauchy surface, the extra two of which
are associated
with the field $W$.
That a Lagrange multiplier is propagating merely signifies that
it is a
determined multiplier, with its evolution derived from the field
equations
\cite{Kelly:1991} and not freely fixable as was done in
\cite{Moffat:1981,Mann:1986} and in the next to last section of
\cite{Kunstatter+Leivo+Savaria:1984} where {\it ad hoc}
constraints were imposed
on the linearized theory in order to obtain the dynamics of a
Kalb-Ramond
theory.
That these constraints cannot exist is clear from the lack of
gauge invariance
in the full NGT action.

The result of vanNieuwenhuizen does however motivate a potential
solution to
this problem, since the massive Kalb-Ramond theory does not
require a conserved
current and yet has no ghost modes, higher order poles or
tachyons.
The additional terms in the action for mNGT (\ref{mNGTAct}) are
introduced in
order to allow the linearized field equations of NGT to take on
this form in the
antisymmetric sector.
These two terms play slightly different roles: the $W^2$ term
causes $W$ to be
determined in terms of metric functions directly ($\alpha$ is
fixed in order to
find the correct form of the kinetic energy terms), and the mass
term for
$g_{[\;]}$ makes the skew sector short-ranged, and ensures that
the linearized
field equations remain consistent when expanding about a flat
background.

Thus mNGT should  have a linearization about Minkowski space of
the form:
\begin{equation}\label{lmNGTflat}
\partial^\alpha F_{\mu\nu\alpha}
+m^2h_{[\mu\nu]}
=J_{[\mu\nu]}.
\end{equation}
The solution to (\ref{lmNGTflat}) can be found by taking a
divergence and
substituting back in to find:
\begin{eqnarray}\label{MKRs}
h_{[\mu\nu]}&=&\square^{-1}\left[J_{[\mu\nu]}
+\case{2}{m^2}\partial^\alpha\partial_{[\mu}J_{[\alpha\nu]]}
\right],\nonumber  \\
\partial^\nu h_{[\mu\nu]}&=&\case{1}{m^2}\partial^\nu
J_{[\mu\nu]}.
\end{eqnarray}
The higher order poles have disappeared, and it can be shown that
the linearized
Hamiltonian is weakly positive definite and that ghost modes are
removed through
the algebraic conditions that couple them locally to the source
in (\ref{MKRs}).
About a more general background one can allow a curvature
coupling term, since
it will not affect the behavior of the fields asymptotically once
the background
is assumed to fall off appropriately.
Choosing the theory that results in this behavior in the
linearized theory will
fix $\alpha$ uniquely.

Returning now to the field equations of mNGT expanded about a GR
background
following from (\ref{mNGTFeq}), one finds:
\begin{equation}\label{minit}
{}^1\! {\cal R}_{\mu\nu}={}^1\! R_{\mu\nu}
+\case{1}{\alpha}\nabla_{[\mu}\nabla^\alpha h_{[\alpha\nu]]}
-\case{1}{2}m^2h_{[\mu\nu]}=T_{\mu\nu}-\case{1}{2}g_{\nu\mu}T,
\end{equation}
where the first order correction to the `Ricci' tensor is given
by:
\begin{equation}\label{1Ricci}
{}^1\! R_{\mu\nu}=
\nabla_\alpha\gamma^\alpha_{\mu\nu}
-\nabla_{(\nu}\gamma^\alpha_{\alpha\mu)}.
\end{equation}
Again ignoring the symmetric GR perturbations, the antisymmetric
part of
(\ref{minit}) is:
\begin{eqnarray}\label{Aspt}
&&\nabla_\alpha\gamma^\alpha_{[\mu\nu]}
+\case{1}{\alpha}\nabla_{[\mu}\nabla^\alpha h_{[\alpha\mu]]}
-\case{1}{2}m^2h_{[\mu\nu]}  \nonumber \\
\quad\quad &=&-\case{1}{2}(\nabla^\alpha
F_{\mu\nu\alpha}+m^2h_{[\mu\nu]})
-2\nabla^\alpha\nabla_{[\mu}h_{[\alpha\nu]]}
+\case{1}{\alpha}
[1+\case{2}{3}\alpha]
\nabla_{[\mu}\nabla^\alpha h_{[\alpha\nu]]}=T_{[\mu\nu]}.
\end{eqnarray}
Requiring that this reduce to the massive Kalb-Ramond field
equations
(\ref{lmNGTflat}) determines the (previously arbitrary) coupling:
$\alpha=3/4$.
The last two terms can be reduced to a curvature term to give:
\begin{equation}\label{equation:KR curved}
\nabla^\alpha F_{\mu\nu\alpha}
+m^2h_{[\mu\nu]}
-4{}^0\!
R_{\;\;[\mu\;\;\nu]}^{\alpha\;\;\beta}h_{[\alpha\beta]}
=-2T_{[\mu\nu]},
\end{equation}
so the skew sector perturbations are well-behaved when perturbing
about any
asymptotically flat GR background.

Expanding the action (\ref{mNGTAct}) to second order (ignoring
surface terms)
gives:
\begin{equation}\label{FL2}
{}^2\! {\cal L}=-{}^2\! R-\case{1}{2}h{}^1\! R
+h^{\mu\nu}{}^1\! R_{\mu\nu}
+h^{\mu\nu}\partial_{[\mu}W_{\nu]}+\case{1}{2}\alpha W^\mu W_\mu
+l^\mu\Gamma_\mu
-\case{1}{4}m^2h^{[\mu\nu]}h_{[\mu\nu]}
\end{equation}
and once compatibility is imposed, followed by the removal of
$W$, this becomes:
\begin{equation}\label{mLNGT2}
{}^2\! {\cal L}=\case{1}{12}F^{\mu\nu\gamma}F_{\mu\nu\gamma}
-\case{1}{4}m^2h^{[\mu\nu]}h_{[\mu\nu]}
-\nabla^\gamma h^{[\mu\nu]}\nabla_\nu h_{[\gamma\mu]}
-(\case{1}{2\alpha}+\case{1}{3})
\nabla_\nu h^{[\mu\nu]}\nabla^\gamma h_{[\mu\gamma]}.
\end{equation}
Choosing $\alpha=3/4$ results in kinetic terms identical to those
of Kalb-Ramond
theory on a GR background, giving the skew sector action:
\begin{equation}\label{mNGTGRB}
{}^2\! {\cal L}=\case{1}{12}F^{\mu\nu\gamma}F_{\mu\nu\gamma}
-\case{1}{4}m^2h^{[\mu\nu]}h_{[\mu\nu]}
-h^{[\mu\nu]}h^{[\alpha\beta]} {}^0\! R_{\alpha\mu\beta\nu},
\end{equation}
which reproduces the linearized field equations (\ref{equation:KR
curved}).
Thus the massive NGT action will be (\ref{mNGTAct}) with
$\alpha=3/4$
\cite{Moffat:1994}, giving the action (\ref{mNGTGRB}) for
perturbations about a
GR background and guaranteeing good asymptotic behavior for these
fields.

Although it has been established that the perturbation equations
about a GR
background are a consistent system resulting in good fall-off for
the skew
sector, it is not clear whether an asymptotic perturbation
actually corresponds
to a global solution (linearization stability).
The (seemingly contrived) asymptotic limit of gauge invariant
kinetic terms
cannot be reflected in the full action, since there is no room
for the
additional gauge invariance in theories constructed from
antisymmetric fields in
this manner.
This means that in general that one expects more (perhaps all 6)
degrees of
freedom in the skew sector evolving as degrees of freedom in a
Cauchy analysis,
whereas in any spacetime that has an asymptotically flat region
only three will
survive.
This situation could be similar to that found in
\cite{Isenberg+Nestor:1977},
where vector fields were seen to increase their degrees of
freedom when
gravitational effects are taken into account.
In order to obtain an asymptotically flat spacetime (with the
reduced degrees of
freedom of the vector fields) from physically reasonable initial
data, the
evolution equations were seen to have to encounter singularities.
This is generally considered to be a sign of instability, and
certainly not a
desirable feature in any theory.
Perturbations about NGT backgrounds should also be considered,
since the
physically interesting NGT solutions are most likely those that
are not `close'
to a GR solution \cite{Cornish+Moffat:1994}.
It is hoped that a more complete analysis of the general
spherically symmetric
system should be able to say something about this issue, since it
will certainly
tell one how many degrees of freedom survive and how they couple
to external
fields, and hopefully something about how the system may or may
not approach an
asymptotically flat spacetime.

It is also true that the form of the action (\ref{mNGTAct}) is
far from unique.
In particular, one could replace the $W^2$ term with some
combination of $W^2$
and $\frac{1}{\sqrt{-g}}g_{(\mu\nu)}\partial_\alpha[{\bf
g}^{[\mu\alpha]}]\partial_\gamma[{\bf g}^{[\nu\gamma]}]$, giving
the same
perturbation equations (\ref{equation:KR curved}), and resulting
in an arbitrary
coupling constant in the action.
Further, since there is nothing preventing one from adding
$\Gamma_{[\;]}$ terms
(they are tensors) or even infinite strings of terms of the form:
$g_{(\alpha\beta)}g^{(\beta\gamma)}\cdots$ or
$g_{[\alpha\beta]}g^{[\beta\gamma]}\cdots$ (which conveniently
disappear in the
asymptotic expansion), there is clearly an infinite number of
actions that do
this.
These examples seem extremely unnatural and will not be
considered further here,
although the results of this paper would not change significantly
for these more
general actions.

\section{Spherically Symmetric Perturbation of the Schwarzschild
solution in a
Coordinate basis}
\label{section:Schw pert}

The absence of a Birkhoff theorem may be derived from the
perturbation equations
(\ref{equation:KR curved}) developed in Section
\ref{section:Massive NGT}.
In general the spherically symmetric fields in the skew sector
will not be
static, although the symmetric sector will remain static in the
perturbation
about the GR solution considered here.
The background metric is Schwarzschild with (coordinate basis)
metric written
as: $g={\rm diag}(A(r),-1/A(r),-r^2,-r^2\sin^2(\theta))$, where
$A(r)=1-2M_s/r$
and $M_s$ is the Schwarzschild mass parameter.
The perturbation considered will be one that is spherically
symmetric but not
necessarily static.
A killing vector analysis yields the general form of the
spherically symmetric
perturbation:
\begin{equation}\label{equation:perturbation metric}
| h_{\mu\nu} |=\left[
\begin{array}{cccc}
h_{00}(t,r)&h_{(01)}(t,r)+h_{[01]}(t,r)&0&0\\
h_{(01)}(t,r)-h_{[01]}(t,r)&h_{11}(t,r)&0&0\\
0&0&h_{22}(t,r)&h_{[23]}(t,r)\sin(\theta)\\
0&0&-h_{[23]}(t,r)\sin(\theta)&h_{22}(t,r)\sin^2(\theta)
\end{array}\right].
\end{equation}
Making a change of coordinates of the background geometry is
equivalent to
making a change of gauge on the perturbation: $\delta
h=\pounds_\varepsilon[g]$,
where
$\varepsilon$ is the spherically symmetric vector gauge parameter
generating
diffeomorphisms between spherically symmetric spacetimes.
This allows one to simplify the form of the perturbation by a
suitable choice of
gauge.
Choosing the gauge parameter as:
\begin{mathletters}
\begin{eqnarray}
\varepsilon^0&=&-\int\left(\frac{h_{(01)}}{A(r)}-\frac{\partial_t
[h_{22}]}{2rA(r)
^2}\right)dr,
 \\
\varepsilon^1&=&\frac{h_{22}}{2r},
\end{eqnarray}
\end{mathletters}
removes the $\theta-\theta$, $\phi-\phi$, and symmetric $t-r$
perturbations altogether, and a remaining gauge transformation
$\varepsilon^0=\epsilon(t)$ allows one to remove an
arbitrary function from the $t-t$ component of the form: $\delta
h_{00}=2A(r)\epsilon(t)$.

The field equations will be written without the source terms for
simplicity
although it is straightforward to include them and relate the
constants of
integration to properties of the source.
First reviewing how the symmetric (in this case identically GR)
perturbations
become static, it is simplest to begin with the field equation:
${}^1\! {\cal R}_{(01)}=0$, which implies:
\begin{equation}
\partial_t[h_{11}(t,r)]=0,
\end{equation}
immediately showing that $h_{11}$ must be static.
By considering ${}^1\! {\cal R}_{22}=0$, it is determined to
be:
\begin{equation}
h_{11}(t,r)=-\frac{2\delta M_s}{rA^2},
\end{equation}
where the integration constant has been combined with $M_s$ and
interpreted as a
perturbation of
the Schwarzschild mass parameter: $\delta M_s$.
Then one considers: $g^{00}\,{}^1\! {\cal
R}_{00}-g^{11}\,{}^1\! {\cal R}_{11}=0$, leading to:
\begin{equation}
h_{00}(t,r)=B(t)A-\frac{2\delta M_s}{r},
\end{equation}
also giving a contribution arising from the perturbed mass
parameter, as well as
an arbitrary function of time as an integration constant,
removable by the
remaining choice of gauge noted above with:
$\epsilon(t)=-B(t)/2$.
Thus one has that the symmetric perturbations are static and
interpretable as
being due to a small change in the total energy of the system:
$\delta M_s$.

In the skew sector, the $t-r$ field equation gives:
\begin{equation}\label{equation:h01}
{}^1\! {\cal
R}_{[01]}=\case{1}{2}(2A^{\prime\prime}-m^2)h_{[01]}=0,
\end{equation}
from which one must conclude that $h_{[01]}$ vanishes outside the
source.
(Primes will denote the derivative of a function of one variable
where
convenient.)
In massless Kalb-Ramond theory, this is the surviving spherically
symmetric
ghost mode which in that case is pure gauge.
When a mass term is added, although these modes are now no longer
pure gauge,
they do not propagate since they are locally coupled to the
source.
It is these modes that one eventually must worry about, since in
the full theory
they may play a nontrivial dynamical role.
In the $\theta-\phi$ sector:
\begin{equation}
{}^1\! {\cal
R}_{[23]}=-\frac{1}{2}\left\{\frac{1}{A}\partial_t^2[h_{[23]}]
-A\partial_r^2[h_{[23]}]
-(A^\prime-\frac{2A}{r})\partial_r[h_{[23]}]
+(\frac{4}{r^2}(1-A)+m^2)h_{[23]}\right\}\sin(\theta)=0.
\end{equation}
In order to derive the asymptotic form of this perturbation, it
is convenient to
define $h_{[23]}=rf(t,r)$,  leading to:
\begin{equation}\label{equation:Schw f}
\frac{1}{A}\partial_t^2[f]-A\partial_r^2[f]
-A^\prime\partial_r[f]
+(\frac{3A^\prime}{r}+\frac{2A}{r^2}+m^2)f=0.
\end{equation}
Introducing the coordinate:
\begin{equation}
r^*=\int\frac{dr}{A(r)}=r+2M_s\ln(\frac{r}{2M_s}-1),
\end{equation}
one obtains (after multiplying by $A$) the partial differential
equation for $f$
in normal form:
\begin{equation}\label{equation:Wave 1}
\partial_t^2[f]-\partial_{r^*}^2[f]
+A(m^2+\frac{2A}{r^2}+\frac{3A^\prime}{r})f=0
\end{equation}
where $r$ is considered as a function of $r^*$ as are $A(r)$ and
$\partial_rA(r)$, and the perturbation $f=f(t,r^*)$.
In this form it is obvious that (\ref{equation:Wave 1}) is a
hyperbolic wave
equation, and that the field $f$ is therefore nonlocally related
to the source.

Using the fact that $1/r-1/r^*\sim o(1/(r^*)^2)$ as
$r^*\rightarrow\infty$, one
keeps only the constant mass term asymptotically, as all other
potential terms
will be dominated by it.
This leaves the massive scalar wave equation to determine the
asymptotic form of
the perturbation:
\begin{equation}\label{equation:Wave equation}
\partial_t^2[f]-\partial_{r^*}^2[f]+m^2f\sim 0.
\end{equation}
The static solution of this is easily seen to have the asymptotic
form:
\begin{equation}
h_{[23]}(r)\sim F_0\frac{r}{m}e^{-mr^*}
\sim F_0\frac{r}{m}e^{-mr} (\frac{r}{2M_s})^{-2mM_s},
\end{equation}
where a factor of $m$ has been introduced in order to make the
constant $F_0$
dimensionless.
The general time dependent case may be handled by noting that the
retarded and
advanced Green functions for the massive scalar wave equation
\cite{Bogoliubov+Shirkov:1959} ($x^2=t^2-\vec{x}^2$):
\begin{equation}\label{equation:Green functions}
D^{{\rm ret, adv}}(x)= D^{{\rm ret, adv}}(t,r)=
\frac{1}{2\pi}\theta(\pm
x^0)\left[\delta(x^2)-\frac{m\theta(x^2)}{2\sqrt{x^2}}J_1(m\sqrt{
x^2})\right],
\end{equation}
depends only on $(t,r)$, and $r^*D^{{\rm ret, adv}}(t,r^*)$ will
solve
(\ref{equation:Wave equation}).
The asymptotic behavior of $h_{[23]}$ is then determined from:
\begin{equation}\label{equation:h23 wave solution}
h_{[23]}\sim rr^* D^{{\rm ret, adv}}(t,r^*).
\end{equation}
Note that the behavior on the light cone is determined from just
the massless
Green function $\delta(x^2)$ \cite{Itzykson+Zuber:1980}, and so
it would appear
that $h_{[23]}$ will behave as $r$ as $r\rightarrow\infty$ along
the forward
light cone.
This is misleading, as it can be demonstrated explicitly
\cite{Reed+Simon:1979}
that for $C^\infty$ initial data with compact spatial support, a
massive
Klein-Gordon field is bounded everywhere by: $\phi\leq d(1+\mid
t\mid)^{-3/2}$,
for some constant $d$, and therefore cannot radiate energy.
This can also be understood by noting that because the field is
massive, the
effects propagating on the light cone must be fields of infinite
energy, and
given some physically reasonable source distribution, these
infinite energy
modes will not be excited.

The existence of time dependent solutions thus proves that
Birkhoff theorem is
not valid in mNGT, although the short-ranged nature of the skew
sector implies
that monopole radiation will not exist.
The symmetric sector has remained static in this system, but as
will be shown in
Section \ref{section:Wyman pert}, through a perturbation about an
approximated
mNGT solution, this will not be the case in general.
The perturbation equations about a mNGT background have not been
given in
covariant form, primarily due to the complication involved
(although it is
possible in principle using a generalization of the inversion of
the
compatibility equation given in \cite{Tonnelat:1982}).
Instead the system may be developed in each case separately, and
the analysis
simplified  by considering the field equations in a vierbein
frame given in the
next Section.

\section{Nonsymmetric Theories in a General Frame}
\label{section:GenBasis intro}

The structure of the compatibility relations and field equations
in nonsymmetric
theories can be formulated in terms of components in a general
moving frame (in
the sense of global section of the general linear frame bundle
${\it
GL}{\cal M}$ of all linear frames over ${\cal M}$).
The formalism given here is essentially a more systematic
development of the
approach in \cite{Vlachynsky:1988}, and differs slightly from
that of Hlavaty
\cite{Hlavaty:1958} in that the (in general nonsymmetric in a
coordinate basis)
connection coefficients have been split up into a connection that
is torsion
free, and another that is purely antisymmetric, instead of
defining two types of
covariant derivative, one associated with the NGT Christoffel
symbols, and
another that is in general non-symmetric and not in general
torsion-free.
The construction here has the advantage of only defining one
covariant
derivative, and the fact that it is torsion-free implies that the
antisymmetric
components in a general (non-coordinate) frame are related in the
standard way
to the structure constants.
In a coordinate basis this is the usual split between the
symmetric and
antisymmetric components, however it is easily generalized to any
basis by
considering the antisymmetric components as a separate
antisymmetric tensor, and
the symmetric components as a torsion-free but generally
non-compatible
connection.

This provides a simple way to split the GR and NGT contribution
in weak field
situations, as well as generating computationally simpler systems
to solve when
inverting the compatibility relations.
Note that although the formalism is developed for a general
basis, the
specialization to a vierbein basis (the reduction of ${\it
GL}{\cal M}$ to
${\it L}{\cal M}$, the Lorentz frame bundle consisting of all
Lorentz frames
above ${\cal M}$) which will be utilized in the rest of this
paper, is
accomplished through the choice of the fiber metric as
$g_{(\;)}\rightarrow\eta$
above all points of the manifold.
This is possible in NGT for the same reason that it is possible
in GR:
mathematically formulating a physical theory in a diffeomorphism
invariant
manner will always allow the introduction of these general linear
frames.
The reduction to Lorentz frames is also possible as one is
assuming that the
symmetric part of the metric that one is attempting to
diagonalize is
nondegenerate, allowing the reduction of the frame bundle.
This construction will be of importance when considering the
canonical analysis
of NGT, as one would like to work in a surface compatible
(generally
non-coordinate) basis in order to avoid specialization to a
particular choice of
time parameter fixed by the foliation of the manifold, and is
easily applied to
other systems with a nonsymmetric metric and connection
\cite{Mann:1986}.

\subsection{Metric, Compatibility and Curvature}
\label{section:GenBasis compat}
The compatibility conditions in a coordinate basis (\ref{Comp2m})
will be
written for convenience as:
\begin{equation}\label{coordcomp}
\partial_\gamma [g_{\mu\nu}]
-g_{\mu\alpha}\Gamma^\alpha_{\nu\gamma}
-g_{\alpha\nu}\Gamma^\alpha_{\gamma\mu}
=-\Delta^0_{\gamma\mu\nu}
\end{equation}
where $\Delta^0$ depends only on the metric or quantities
directly derivable
from it (and possibly other quantities, but for the purposes of
this
construction it does not depend on the connection coefficients).
Parallel transport (and the related covariant derivative) will
then be defined
using just the symmetric part of the coordinate basis connection,
and its action
on the (coordinate) basis vectors is:
\begin{equation}\label{parallaltransp}
\nabla_{e_\alpha}[e]_\beta=\Gamma^\gamma_{(\alpha\beta)}e_\gamma,
\quad
\nabla_{e_\alpha}[\theta]^\gamma=-\Gamma^\gamma_{(\alpha\beta)}
\theta^\beta,
\end{equation}
and the connection is split into a symmetric connection and an
antisymmetric
tensor:
\begin{equation}\label{Connectionsplit}
\Gamma^\gamma_{\mu\nu}\rightarrow\Gamma^\gamma_{(\mu\nu)}
+\Lambda^\gamma_{[\mu\nu]}.
\end{equation}
Thus $\Gamma$ will refer from this point onwards to the
torsion-free (symmetric
in a coordinate basis) part of the connection, and $\Lambda$ to
the remaining
tensor contribution.
In this way, $\Gamma$ is a torsion-free (but non-compatible)
covariant
derivative since:
\begin{equation}\label{coordtorsion}
T^\gamma_{\mu\nu}
=\theta^\gamma\left[\nabla_{e_\mu}e_\nu-\nabla_{e_\nu}e_\mu
-[e_\mu,e_\nu]\right]
=2\Gamma^\gamma_{[\mu\nu]}=0.
\end{equation}
The compatibility equation (\ref{coordcomp}) then becomes:
\begin{equation}\label{compat}
\nabla_{e_\gamma}[g]_{\mu\nu}
=e_\gamma[g_{\mu\nu}]
-g_{\mu\alpha}\Gamma^\alpha_{\gamma\nu}
-g_{\alpha\nu}\Gamma^\alpha_{\gamma\mu}
=g_{\mu\alpha}\Lambda^\alpha_{\nu\gamma}
+g_{\alpha\nu}\Lambda^\alpha_{\gamma\mu}
-\Delta^0_{\gamma\mu\nu},
\end{equation}
where the basis vectors are just directional derivatives along
the coordinates:
$e_\gamma[\;]=\partial_\gamma[\;]$.

With this definition of the covariant derivative and related
connection
coefficients, the geometric curvature is found as usual from:
\begin{eqnarray}\label{Riemann}
R^\alpha_{\;\beta\mu\nu}
&=&\theta^\alpha[(\nabla_{e_\mu}\nabla_{e_\nu}
-\nabla_{e_\nu}\nabla_{e_\mu}-
\nabla_{[e_\mu,e_\nu]})e_\beta] \nonumber \\
&=&e_\mu[\Gamma^\alpha_{\nu\beta}]
-e_\nu[\Gamma^\alpha_{\mu\beta}]+
\Gamma^\gamma_{\nu\beta}\Gamma^\alpha_{\mu\gamma}
-\Gamma^\gamma_{\mu\beta}\Gamma^\alpha_{\nu\gamma},
\end{eqnarray}
and defining the two independent contractions:
\begin{mathletters}
\begin{eqnarray}\label{Riccis}
R^1_{\mu\nu}&=&R^\alpha_{\;\mu\alpha\nu}
=e_\alpha[\Gamma^\alpha_{\nu\mu}]
-e_\nu[\Gamma^\alpha_{\alpha\mu}]+
\Gamma^\gamma_{\nu\mu}\Gamma^\alpha_{\alpha\gamma}
-\Gamma^\gamma_{\alpha\mu}\Gamma^\alpha_{\nu\gamma},  \\
R^2_{\mu\nu}&=&R^\alpha_{\;\alpha\mu\nu}
=e_\mu[\Gamma^\alpha_{\nu\alpha}]
-e_\nu[\Gamma^\alpha_{\mu\alpha}],
\end{eqnarray}
\end{mathletters}
The Ricci tensor will be defined as:
\begin{eqnarray}\label{Ricci}
R_{\mu\nu}&=&
R^1_{\mu\nu}-\case{1}{2}R^2_{\mu\nu}\nonumber   \\
&=&e_\alpha[\Gamma^\alpha_{\nu\mu}]-\case{1}{2}
(e_\nu[\Gamma^\alpha_{\alpha\mu}]
+e_\mu[\Gamma^\alpha_{\nu\alpha}])
+\Gamma^\gamma_{\nu\mu}\Gamma^\alpha_{\alpha\gamma}
-\Gamma^\gamma_{\alpha\mu}\Gamma^\alpha_{\nu\gamma}.
\end{eqnarray}
This particular combination is symmetric, and obviously reduces
to the GR Ricci
tensor when the NGT antisymmetric terms vanish.
Decomposition of (\ref{NGTRicci}) into $R_{\mu\nu}$ and another
that depends on
$\Lambda$ as:
$R^{\text{\tiny{NS}}}_{\mu\nu}=R_{\mu\nu}+R^\Lambda_{\mu\nu}$
where ($\Lambda_\mu=\Lambda^\alpha_{\mu\alpha}$) gives:
\begin{equation}
R^\Lambda_{\mu\nu}=\nabla_{e_\alpha}[\Lambda]^\alpha_{\mu\nu}
+\nabla_{e_{[\mu}}[\Lambda]_{\nu]}
+\Lambda^\alpha_{\mu\beta}\Lambda^\beta_{\nu\alpha}.
\end{equation}

A more general basis is introduced at each point on the manifold
through
$e_A=E_A^{\;\;\mu}e_\mu$, where $E$ is locally an element of
$Gl(4,{\cal
R})$, and these bases are smoothly joined up to form sections of
the tangent
bundle $T({\cal M})$ \cite{Choquet-Bruhat+:1989,Nakahara:1990}.
The general basis vectors are then given in terms of a coordinate
basis through
the vierbein-like quantities, which can be used to translate
tensors from one
choice of basis to the other:
\begin{equation}\label{Evbns}
e_A=E_A^{\;\;\mu}e_\mu, \quad
E_A^{\;\;\mu}E_B^{\;\;\nu}g_{\mu\nu}=g_{AB}, \quad {\rm etc.}.
\end{equation}
(In the usual orthonormal basis, one transforms the symmetric
part of the metric
to the Minkowski space metric, and the $E$'s provide the
isomorphism between
coordinate basis tensors and locally Lorentzian tensors.)
The dual basis of $T^*({\cal M})$ is introduced through the usual
relation:
$\theta^A[e_B]=\delta^A_B$, and the inverse of the vierbeins is
defined through:
$E_A^{\;\;\mu}E^A_{\;\;\nu}=\delta^\mu_\nu$.
In this paper capital letters from the beginning of the alphabet:
$A,B,C,\cdots$
will refer to components of the object decomposed in the general
basis.

Parallel transport of the basis vectors now defines the
generalized connection
coefficients:
\begin{equation}\label{paralltptncb}
\nabla_{e_A}[e]_B=\Gamma^C_{AB}e_C,\quad
\nabla_{e_A}[\theta]^C=-\Gamma^C_{AB}\theta^B.
\end{equation}
The definition of the basis in (\ref{Evbns}) implies that it is
no longer a
coordinate basis in general, and hence the directional
derivatives no longer
necessarily commute, giving rise to the structure constants:
\begin{mathletters}
\begin{equation}\label{ncbStructure}
[e_A,e_B]={C_{AB}}^Ce_C,
\end{equation}
given by:
\begin{equation}\label{Stcnst}
{C_{AB}}^C=E^C_{\;\;\nu}(
E_ A^{\;\;\mu}\partial_\mu[E_ B^{\;\;\nu}]
-E_ B^{\;\;\mu}\partial_\mu[E_ A^{\;\;\nu}]),
\end{equation}
\end{mathletters}
calculated by noting that $e_A[\;]={E_A}^\mu\partial_\mu[\;]$.
This also implies that a torsion-free connection will no longer
be symmetric,
and vanishing torsion now gives:
\begin{equation}\label{ncbtorsion}
T^A_{BC}
=\theta^A\left[\nabla_{e_B}e_C-\nabla_{e_C}e_B-[e_B,e_C]\right]
=2\Gamma^A_{[BC]}-{C_{BC}}^A=0,
\end{equation}
allowing one to determine the antisymmetric part as usual from
the structure
constants.
This is the motivation for splitting up the connection in this
way.
Given some alternate split where $\Gamma$ is not torsion free,
one would have to
distinguish between the effects of the general basis on the skew
part of the
connection coefficients, and that of the NGT effects (themselves
tensors).

The compatibility condition (\ref{coordcomp}) can now be written
as:
\begin{equation}\label{ncbcompat}
\nabla_{e_C}[g]_{AB}
=g_{AD}\Lambda^D_{BC}
+g_{DB}\Lambda^D_{CA}
-\Delta^0_{CAB},
\end{equation}
where since $\Delta^0$ and $\Lambda$ are tensors, they are just
redefined by
multiplication by the appropriate combination of vierbeins.
The symmetric part of this can now be solved for the symmetric
part of $\Gamma$
in terms of the antisymmetric part, the structure constants, and
$\Lambda$, to
give:
\begin{equation}\label{symmconn}
\Gamma_{C(AB)}=\case{1}{2}\Delta_{C(AB)}-\Gamma_{A[BC]}+\Gamma_{B
[CA]}
-A^D_{\;A}\Lambda_{DBC}+A^D_{\;B}\Lambda_{DCA},
\end{equation}
where the quantities:
\begin{equation}\label{defns}
\Gamma_{ABC}=g_{(AD)}\Gamma^D_{BC},\quad
\Lambda_{ABC}=g_{(AD)}\Lambda^D_{BC},\quad
A^A_{\;B}=S^{(AC)}g_{[CB]},
\end{equation}
have been defined for convenience, and $S$ is the inverse of the
symmetric part
of the metric defined by: $S^{(AB)}g_{(BC)}=\delta^A_C$.
Also appearing is the symmetric part of:
\begin{equation}\label{delta}
\Delta_{CAB}=e_B[g_{CA}]+e_A[g_{BC}]-e_C[g_{AB}]
+\Delta^0_{BCA}+\Delta^0_{ABC}-\Delta^0_{CAB}.
\end{equation}
The antisymmetric part of the compatibility conditions can now be
recast (using
(\ref{symmconn})) as 24 algebraic equations for $\Lambda$:
\begin{equation}\label{compantisymm}
\Lambda_{CAB}-A^D_{\;A}A^E_{\;B}\Lambda_{ECD}
-A^D_{\;A}A^E_{\;C}\Lambda_{EBD}
+A^D_{\;B}A^E_{\;A}\Lambda_{ECD}
+A^D_{\;B}A^E_{\;C}\Lambda_{EAD}
=\Omega_{C[AB]},
\end{equation}
where:
\begin{eqnarray}\label{omega}
\Omega_{C[AB]}&=&\case{1}{2}(\Delta_{C[AB]}
+A^D_{\;B}\Delta_{D(CA)}
-A^D_{\;A}\Delta_{D(BC)}) \nonumber \\
\quad\quad &+&A^D_{\;A}(\Gamma_{B[CD]}+\Gamma_{C[BD]})
-A^D_{\;B}(\Gamma_{A[CD]}+\Gamma_{C[AD]})
-A^D_{\;C}\Gamma_{D[AB]}.
\end{eqnarray}

The method for solving the compatibility conditions is to first
determine the
auxiliary quantities appearing in this relation:
($A,\Gamma_{[\;]},\Delta,\Omega$) in terms of the vierbeins and
metric
quantities, then solve for $\Lambda$ through
(\ref{compantisymm}), determine
$\Gamma_{(\;)}$ from (\ref{symmconn}), and then use $S$ with
$\Gamma_{[\;]}$ and
$\Gamma_{(\;)}$ to form $\Gamma^A_{BC}$ and $\Lambda^A_{BC}$.
This may not seem like much of a simplification, but when
specialized to a
Lorentz frame, many of these quantities simplify considerably (as
is the case in
the Wyman sector in Section \ref{section:Wyman}).

The curvature tensor (\ref{Riemann}) becomes:
\begin{eqnarray}\label{ncbRiemann}
R^A_{\;BCD}
&=&\theta^A[(\nabla_{e_C}\nabla_{e_D}
-\nabla_{e_D}\nabla_{e_C}-
\nabla_{[e_C,e_D]})e_B] \nonumber \\
&=&e_C[\Gamma^A_{DB}]
-e_D[\Gamma^A_{CB}]+
\Gamma^E_{DB}\Gamma^A_{CE}
-\Gamma^E_{CB}\Gamma^A_{DE}
-{C_{CD}}^E\Gamma^A_{EB},
\end{eqnarray}
and the contractions:
\begin{mathletters}
\begin{eqnarray}
R^1_{AB}&=&R^C_{\;ACB}=e_C[\Gamma^C_{BA}]-e_B[\Gamma^C_{CA}]
+\Gamma^D_{BA}\Gamma^C_{CD}-\Gamma^D_{CA}\Gamma^C_{BD}-{C_{CB}}^D
\Gamma^C_{DA},
\\
R^2_{AB}&=&R^C_{\;CAB}=e_A[\Gamma^C_{BC}]-e_B[\Gamma^C_{AC}]-{C_{
AB}}^D\Gamma^C_{
DC},
\end{eqnarray}
\end{mathletters}
combine to give the Ricci tensor:
\begin{eqnarray}\label{ncbRicci}
R_{AB}&=&
R^1_{AB}-\case{1}{2}R^2_{AB} \nonumber \\
&=&e_C[\Gamma^C_{BA}]
-e_B[\Gamma^C_{CA}]
-\case{1}{2}e_A[\Gamma^C_{BC}]
+\case{1}{2}e_B[\Gamma^C_{AC}]\nonumber  \\
&&\quad+\Gamma^D_{BA}\Gamma^C_{CD}
-\Gamma^D_{CA}\Gamma^C_{BD}
-{C_{CB}}^D\Gamma^C_{DA}
+\case{1}{2}{C_{AB}}^D\Gamma^C_{DC}\nonumber  \\
&=&e_C[\Gamma^C_{BA}]
-e_B[\Gamma^C_{CA}]
-\case{1}{2}e_A[\Gamma^C_{BC}]
+\case{1}{2}e_B[\Gamma^C_{AC}] \nonumber \\
&&\quad+\Gamma^D_{BA}\Gamma^C_{CD}
-\Gamma^D_{CB}\Gamma^C_{DA}
+\Gamma^D_{[AB]}\Gamma^C_{DC}.
\end{eqnarray}
In the split $R_{AB}^{\text{\tiny{NS}}}=R_{AB}+R^\Lambda_{AB}$:
\begin{equation}\label{ncbRNgt}
R^\Lambda_{AB}=
\nabla_{e_C}[\Lambda]^C_{AB}
+\nabla_{e_{[A}}[\Lambda]_{B]}
+\Lambda^C_{AD}\Lambda^D_{BC}
\end{equation}
as expected.

Since by construction $\Gamma$ is a torsion free connection, and
(\ref{ncbRiemann}) is the standard curvature tensor constructed
from it, one
obtains the usual Bianchi identities \cite{Nakahara:1990} on the
curvature
tensor.
One should note though that the connection is not compatible, and
so the
rotation coefficients are not antisymmetric.
The relevant Ricci tensors are also not constructed in the same
manner as in GR,
so the implications of these identities are somewhat different.
The first Bianchi identity gives the usual cyclic identity on the
last three
indices of (\ref{ncbRiemann}), and leads to the result:
\begin{equation}
R^1_{[AB]}=\case{1}{2}R^2_{AB}
\end{equation}
when one contracts on any lowered index.
(This can also be proven directly using the Jacobi identity.)
This tells us that the NGT Ricci tensor is symmetric
($R_{[AB]}=0$) in general,
not just in a coordinate basis.

A detailed study of the contractions of the second Bianchi
identity (the cyclic
covariant derivative):
\begin{equation}\label{equation:2nd Bianchi}
\nabla_{e_C}[R]^A_{\;BDE}+\nabla_{e_D}[R]^A_{\;BEC}+\nabla_{e_E}[
R]^A_{\;BCD}=0,
\end{equation}
should result in a derivation of the equations of motion for
matter fields
\cite{Moffat:1987,Legare+Moffat:1995} from the field equations.

\subsection{The NGT Action and Field Equations in a General
Basis}
\label{FEQ}

The translation of the field equations
(\ref{mVars:b},\ref{Comp2m},\ref{mNGTFeq}) is accomplished
through an almost
straightforward substitution:
\begin{mathletters}
\label{mNGTF}
\begin{eqnarray}
\label{mNGTF:b}
\Lambda_A&=&0, \\
\label{mNGTF:c}
\nabla_{e_B}[{\bf g}]^{[AB]}&=&\alpha{\bf g}^{(AB)}W_B,\\
\label{mNGTF:a}
{\cal
R}_{AB}:=R^{\text{\tiny{NS}}}_{AB}+\nabla_{e_{[A}}[W]_{_{B]}}
&-&\case{1}{2}\alpha
W_{A}W_{B}
-\case{1}{4}m^2M_{AB}
=0,
\end{eqnarray}
\end{mathletters}
where the density is $\sqrt{-g}=\sqrt{-{\rm det}(g_{AB})}$, the
mass tensor:
\begin{equation}\label{equation:mass tensor}
M_{AB}=g_{[AB]}-g_{CA}g_{BD}g^{[CD]}
+\case{1}{2}g_{BA}g^{[CD]}g_{[CD]}
\end{equation}
has been defined, and the tensor appearing in the compatibility
equations is:
\begin{equation}\label{deltas}
\Delta^0_{CAB}=-\case{2}{3}\alpha
(g_{A[C}g_{D]B}+\case{1}{2}g_{AB}g_{[CD]})g^{(DE)}W_E.
\end{equation}
One must be careful to treat totally antisymmetric derivatives
properly (the
structure constants now come into the curl of a vector), and
translate the
metric density properly.

In order to define the action, one should note that the inverse
of the metric is
now: $g_{AB}g^{BC}= g^{CB}g_{BA}=\delta^C_A$, and the direct
translation of the
density results in:
$\sqrt{-g}\rightarrow\sqrt{-EgE^t}=E\sqrt{-g}$ where $g={\rm
det}(g_{AB})$ and $E={\rm det}(E^A_{\;\;\mu})$.
Then (\ref{mNGTAct}) is rewritten:
\begin{equation}\label{mNGTActncb}
S=\int_{\cal M}d^4x\,E\left\{-{\bf
g}^{AB}R^{\text{\tiny{NS}}}_{AB}
-{\bf g}^{AB}\nabla_{e_{[A}}[W]_{_{B]}}+{\bf l}^A\Lambda_A
+\case{1}{2}\alpha{\bf g}^{(AB)}W_AW_B
+\case{1}{4}m^2{\bf g}^{[AB]}g_{[AB]}\right\}.
\end{equation}
(Note that in a Lorentz basis, the inverse of the metric is not
$\eta$.)

Deriving the equations of motion from this action should be
approached with
care.
As it stands there are too many fields (the metric and the
vierbeins share
degrees of freedom) and one typically must choose either a
coordinate basis (as
in Section \ref{section:Massive NGT}), a Lorentz basis (so that
all symmetric
metric degrees of freedom are contained in the vierbeins), or a
well-defined
combination of the two.
One must also realize that (\ref{mNGTActncb}) as it stands
assumes that the
connection $\Gamma$ is torsion-free {\it a priori}, so that when
varying the
vierbein, $\Gamma_{[\;]}$ must be varied as well.
As an alternative, one may impose the torsion-free condition
through additional
Lagrange multiplier terms: $L^{BC}_AT^A_{BC}$ in the action,
varying the full
connection coefficients and vierbeins separately.

\section{Approximation of the Wyman Sector Solution in a Vierbein
Basis}
\label{section:Wyman}

In general, the spherically symmetric Killing vector analysis for
a $(0,2)$
tensor gives both $t-r$ and $\theta-\phi$ skew components.
However it is possible to show from the general spherically
symmetric field
equations that it is consistent to put either (or both) of these
skew components
to zero separately, since in either case one loses the
corresponding field
equation, and the system of equations remains consistent.
Whether it is physically reasonable to do this or not depends on
the details of
the matter coupling in the theory, and how it alters the global
behavior of the
skew sector.
Here will be considered the field equations for what will be
referred to as the
Wyman sector \cite{Wyman:1950} (keeping just the $\theta-\phi$
sector), although
the asymptotics of the $t-r$ sector will be discussed briefly at
the end of this
section, where it will be argued that there are no static
solutions with
asymptotic behavior that is dominated by Schwarzschild (or
equivalently,
Newtonian) effects.
This will allow an analysis of the perturbation equations for the
spherically
symmetric modes, in order to see the effects of the antisymmetric
background.

In a coordinate basis, the Wyman metric looks like:
\begin{equation}
| g_{\mu\nu}| ={\rm
diag}\{\gamma(r),-\alpha(r),-r^2,-r^2\sin^2(\theta)\},\quad
g_{[23]}=f(r)\sin(\theta).
\end{equation}
(An appropriate coordinate system has been chosen in order to
remove
the symmetric $t-r$ metric component, and fix the $\theta-\theta$
component.)
Introducing the usual choice of vierbein (using the functions
defined by:
$F=f(r)/r^2, E_0=1/\sqrt{\gamma(r)}, E_1=1/\sqrt{\alpha(r)}$):
\begin{equation}\label{equation:vierbeins}
E_A^{\;\;\mu}={\rm diag}\{E_0^{\;\;0}=E_0,\;
E_1^{\;\;1}=E_1,\;
E_2^{\;\;2}=\frac{1}{r},\;
E_3^{\;\;3}=\frac{1}{r\sin(\theta)}\},
\end{equation}
the metric becomes:
\begin{equation}\label{equation:Wyman metric}
| g_{(AB)} |=\eta_{AB},\quad
g_{[23]}=F,
\end{equation}
and the density $\sqrt{-g}=\sqrt{1+F^2}$.

At this point one can invert the compatibility conditions and
compute the field
equations using the method of Section \ref{section:GenBasis
intro}, given in
some detail in Appendix \ref{appendix:Wyman}.
No attempt will be made here to solve the field equations
exactly, although
numerical evidence for the existence of an exact solution with
asymptotic
behavior that matches that given here has been found
\cite{Cornish:1994},
ensuring that the approximations given come from a global
solution.
Instead, an approximation will be given that describes the
asymptotic behavior
of the exact solution.
The idea will be to consider the skew sector as a small
correction (of order
some small dimensionless parameter $\kappa$, to be explicitly
defined later) to
the Schwarzschild solution far enough away from the source.
This should be reasonable since one expects from the results of
the perturbation
in Section \ref{section:Schw pert} that the skew sector will
behave
asymptotically as a decaying exponential, while the symmetric
sector should
behave as $\sim 1/r$, so that far enough away from the
gravitational source the
skew sector should be completely dominated by GR effects.

To lowest order in $\kappa$ (the skew sector) the work is already
done, as the
field equation for the skew function will be essentially the same
as the static
perturbation about a Schwarzschild background already considered
in Section
\ref{section:Schw pert}.
In the vierbein basis, this is derived as before from ${\cal
R}_{[23]}$
(\ref{equation:Wyman field equations}), and gives:
\begin{equation}\label{equation:Wyman F}
A\partial_r^2[F]+(A^\prime+\frac{2A}{r})\partial_r[F]-\frac{2}{r}
(A^\prime+\frac
{A}{r})F-m^2F=0,
\end{equation}
and it is trivial to see that when using: $F=f/r$, this reduces
to the static
limit of (\ref{equation:Schw f}), giving the asymptotic form for
$F$:
\begin{equation}\label{equation:Wyman asymptF}
F\sim F_0\frac{e^{-mr^*}}{mr}.
\end{equation}

One now must consider how the presence of the skew sector affects
the symmetric
sector, particularly whether it really is a higher order effect.
The asymptotic form of these corrections due to $F$ may be
calculated by
considering order $\kappa^2$ corrections to the vierbeins (order
$\kappa$ terms
will not depend on $F$, and so will be solely $\delta M_s$
corrections),
calculated from the symmetric field equations with $F$ from
(\ref{equation:Wyman
asymptF}) acting as a source.
Writing the corrections to the vierbeins as: $E_{0,1}\rightarrow
E_{0,1}+E_{0,1}^{(2)}$, and the corrections to the field
equations as ${\cal
R}^{(2)}$, one calculates:
\begin{equation}\label{equation:Wyman appx12}
{\cal R}_{00}^{(2)}+{\cal
R}_{11}^{(2)}=-\frac{2A}{r}\partial_r[\sqrt{A}E^{(2)}_0+\frac{E^{
(2)}_1}{\sqrt{A
}}]
-AFF^{\prime\prime}-\frac{3}{2}A(F^\prime)^2
-\frac{2A}{r}FF^\prime=0,
\end{equation}
which, after translating it into a differential equation in $r^*$
and keeping
only the asymptotically dominant terms, results in:
\begin{equation}\label{equation:Wyman asympt12}
\partial_{r^*}[\sqrt{A}E^{(2)}_0+\frac{E^{(2)}_1}{\sqrt{A}}]
\sim -\frac{5}{4}(F_0)^2\frac{e^{-2mr^*}}{r^*}.
\end{equation}
this integrates to give (the constant of integration is ignored
as one could
eliminate it through an appropriate choice of gauge as in Section
\ref{section:Schw pert}):
\begin{equation}\label{equation:Wyman asymptsum}
\sqrt{A}E^{(2)}_0+\frac{E^{(2)}_1}{\sqrt{A}}\sim
\frac{5}{4}(F_0)^2\frac{e^{-2mr^*}}{2mr^*}.
\end{equation}
Considering next ${\cal R}_{33}^{(2)}$ and using
(\ref{equation:Wyman
appx12}) leads to the asymptotic equation:
\begin{equation}\label{equation:Wyman asympt1}
\partial_{r^*}[r\sqrt{A}E^{(2)}_1]\sim-\case{5}{4}(F_0)^2e^{-2mr^
*}.
\end{equation}
The solution of this combined with the results of
(\ref{equation:Wyman
asymptsum}) gives (once again the constant of integration is
ignored, this time
as it would be interpretable as a perturbation of the mass
parameter and not due
to the effects of the skew sector):
\begin{equation}\label{equation:Wymna d1}
E^{(2)}_1\sim \frac{5}{4}(F_0)^2\frac{e^{-2mr^*}}{2mr^*},\quad
E^{(2)}_0\sim o(\frac{e^{-2mr^*}}{(r^*)^2}),
\end{equation}
where the dominant correction to the symmetric sector is
$E^{(2)}_1$, and
$E^{(2)}_0$ is down by $o(1/r^*)$.
It is not hard to see that these corrections are indeed of an
order higher than
the effects in the skew sector.
Clearly one may define a small parameter
$\kappa=F_0\exp(-mr^*_0)$, where
$r^*_0$ is chosen such that $F(r^*_0)\ll 1$, to define the small
size of the
skew sector when $r^*>r^*_0$.
The corrections to the symmetric sector are seen to be of order
$\kappa^2$, and
will therefore be neglected in the approximation of the
background required in
Section \ref{section:Wyman pert}.

One may attempt to do the same sort of analysis keeping the
$g_{[01]}$
component, however the linearized field equation implies
immediately that the
field must vanish (it is identical to (\ref{equation:h01})).
Considering higher orders in the field in an attempt to generate
a solution
other than this trivial result, the third order correction gives
(writing
$W(r)=\sqrt{X(r)}$):
\begin{equation}\label{equation:Wyman W3}
{\cal
R}_{[01]}=-\frac{\sqrt{X}}{6}\left[A\partial_r^2[X]
+(A^\prime+\frac{2A}{r})\partial_r[X]
+(\frac{12A}{r^2}+\frac{4A^\prime}{r}+\frac{3}{2}m^2)X+\frac{12A^
\prime}{r}+3m^2
\right]=0.
\end{equation}
Writing $X=Y/r$ and transforming to the $r^*$ coordinate as
before gives in
canonical form:
\begin{equation}\label{equation:W3 canonical}
\partial^2_{r^*}[Y]+A(\frac{12A}{r^2}+\frac{3A^\prime}{r}+\frac{3
}{2}m^2)Y+A(12A
^\prime+3m^2r)=0,
\end{equation}
and keeping the dominant terms:
\begin{equation}\label{equation:W3 canonical asyptotic}
\partial^2_{r^*}[Y]+\case{3}{2}m^2Y+3m^2r^*=0,
\end{equation}
easily giving the asymptotic form of the solution:
\begin{equation}\label{equation:W3 soln}
W^2(r^*)=-2+\frac{a}{r^*}\cos(\sqrt{\case{3}{2}}mr^*)
+\frac{b}{r^*}\sin(\sqrt{\case{3}{2}}mr^*),
\end{equation}
(where $(a,b)$ are arbitrary constants).
The dominant part of this solution implies that $W$ is imaginary,
and must be
discarded.
However one also sees that this solution is not in fact a small
correction to
the Schwarzschild metric asymptotically, and would have to be
discarded for that
reason alone.
This is not surprising as one is trying to match a function that
is small
asymptotically (by hypothesis) to one that is constant, so
keeping higher orders
in $W$ will not change this.
This implies that nontrivial static solutions that include this
sector fail to
be dominantly Schwarzschild for large $r$.
This of course does not exclude solutions with asymptotic
behavior that is of
some other form, nor can one exclude the possibility that $W$ is
nonvanishing
only inside some finite radius.

\section{Spherically Symmetric Perturbation about a Wyman
Background}
\label{section:Wyman pert}

In an attempt to consider the perturbation equations for NGT
about a general
non-symmetric background, one finds that the compatibility
conditions prevent
one from formulating the inversion in a useful form.
This means that a fairly straightforward covariant formulation
(like that given
in Section \ref{section:Massive NGT}) is not feasible, and
instead one must
treat each situation separately, in this case a spherically
symmetric
perturbation about the approximated mNGT Wyman solution given in
the previous
section.
Here it is demonstrated that despite the remaining gauge freedom
in the
symmetric sector, both symmetric functions will in general pick
up time
dependence from the skew sector.
Although this cross coupling is demonstrated explicitly in a
perturbative
scenario, it will certainly persist in a more general sense.
The results here will show that the perturbations in the
symmetric sector pick
up time dependence that is algebraically determined by the skew
function $F$,
without themselves becoming independent degrees of freedom.
The canonical analysis of the general spherically symmetric
system will address
rigorously how many degrees of freedom exist in each sector.
If there are more in the nonperturbative theory, one can examine
the dynamical
approach to an asymptotically flat spacetime
looking for possible singular behavior similar to that found in
\cite{Isenberg+Nestor:1977}.

The perturbation of the Wyman metric (\ref{equation:Wyman
metric}) in a
coordinate basis will look identical to
(\ref{equation:perturbation metric})
(using the gauge choice to simplify it as before).
The background vierbeins will be the same as those in the Wyman
solution
(\ref{equation:vierbeins}), where now the perturbations of the
vierbeins and
skew metric functions are related to perturbations in the
coordinate basis by:
\begin{equation}
\delta W=\frac{h_{[01]}(t,r)}{\sqrt{\alpha(r)\gamma(r)}},\quad
\delta F=\frac{\delta h_{[23]}(t,r)}{r^2},\quad
\delta
E_0=-\frac{1}{2}\frac{h_{00}(t,r)}{\gamma(r)^{\frac{3}{2}}},\quad
\delta
E_1=-\frac{1}{2}\frac{h_{11}(t,r)}{\alpha(r)^{\frac{3}{2}}}.
\end{equation}
In the vierbein basis the metric perturbation has nonvanishing
components:
\begin{equation}
h_{[01]}=\delta W,\quad h_{[23]}=\delta F.
\end{equation}
The approximation of the background Wyman solution given in
Section
\ref{section:Wyman} greatly simplifies the algebra necessary to
develop the
perturbation given in Appendix \ref{appendix:Wyman perturbation}.
Approximating the symmetric sector by the Schwarzschild solution
and the
antisymmetric sector by (\ref{equation:Wyman asymptF}), first
order in this
static antisymmetric background is kept, as is the first order in
the
perturbations.
As one shall see, this will be a reasonable approximation since
it will be
possible to keep the perturbations small compared to the
background by an
appropriate choice of integration constants (similarly to $\delta
M_s$ in the
Schwarzschild case).

The field equation: ${}^1\! {\cal R}_{[01]}=0$, yields
precisely
the same field equation as in the Schwarzschild case
(\ref{equation:h01}),
allowing one to immediately set $\delta W=0$.
The symmetric part: ${}^1\! {\cal R}_{(01)}=0$, can be written
as
a total time derivative:
\begin{equation}
{}^1\! {\cal
R}_{(01)}=\partial_t\left[\frac{2}{r\sqrt{A}}\delta
E_1
+\frac{3}{2}F^\prime\delta F
+(\frac{1}{r}-\frac{A^\prime}{2A})F\delta F
+F\partial_r[\delta F]\right]=0.
\end{equation}
This last field equation is then integrated, introducing an
arbitrary static
function $\delta E(r)$:
\begin{equation}\label{equation:Wyman deltaE1}
\frac{2}{r\sqrt{A}}(\delta E_1-\delta E)
=-\frac{3}{2}F^\prime\delta F
-(\frac{1}{r}-\frac{A^\prime}{2A})F\delta F
-F\partial_r[\delta F].
\end{equation}
Now computing
(${\rm Tr}[{\cal R}_{AB}]:={\cal R}_{00}+{\cal R}_{11}+2{\cal R}_{22}$):
\begin{eqnarray}
{\rm Tr}[{}^1\! {\cal
R}_{AB}]&=&\frac{4}{r^2}\partial_r[r\sqrt{A}\delta E_1]
-\frac{F}{A}\partial_t^2[\delta F]
+3AF\partial_r^2[\delta F] \nonumber \\
&+&(\frac{8A}{r}F+2A^\prime F+5A F^\prime)\partial_r[\delta F]
+(3AF^{\prime\prime}+\frac{8A}{r}F^\prime+2A^\prime
F^\prime-m^2F)\delta F=0,
\end{eqnarray}
and inserting (\ref{equation:Wyman deltaE1}) gives:
\begin{eqnarray}\label{equation:Wyman TrRsimp}
{\rm Tr}[{}^1\! {\cal
R}_{AB}]&=&\frac{4}{r^2}\partial_r[r\sqrt{A}\delta E] \nonumber
\\
&-&F\left( \frac{1}{A}\partial_t^2[\delta F]
-A\partial_r^2[\delta F]
-(A^\prime+\frac{2A}{r})\partial_r[\delta F]
+(\frac{2A^\prime}{r}+\frac{2A}{r^2}+m^2)F\delta F\right)=0.
\end{eqnarray}
Also,
\begin{eqnarray}\label{equation:Wyman R00plusR11}
{}^1\! {\cal R}_{00}+{}^1\! {\cal R}_{11}
&=&\frac{2A}{r}\partial_r[\sqrt{A}\delta E_0+\frac{\delta
E_1}{\sqrt{A}}]\nonumber  \\
&+&\frac{F}{A}\partial_t^2[\delta F]
+AF\partial_r^2[\delta F]
+A(\frac{2F}{r}+3F^\prime)\partial_r[\delta F]
+A(\frac{2F^\prime}{r}+F^{\prime\prime})\delta F=0,
\end{eqnarray}
gives the equality of spatial derivatives of $\sqrt{A}\delta E_0$
and $\delta
E_1/\sqrt{A}$ up to order $F$.
This will be useful when considering:
\begin{eqnarray}\label{equation:R23 one}
{}^1\! {\cal R}_{[23]}
&=&(\frac{F}{r}-\frac{F^\prime}{2})A\partial_r\left[\frac{\delta
E_1}{\sqrt{A}}-\sqrt{A}\delta E_0\right]
-\left[AF^{\prime\prime}+(A^\prime+\frac{2A}{r})F^\prime
-\frac{2}{r}(A^\prime+\frac{A}{r})F\right]\frac{\delta
E_1}{\sqrt{A}}\nonumber \\
&&\quad+\frac{1}{2}\left(\frac{1}{A}\partial_t^2[\delta F]
-A\partial_r^2[\delta F]
-(A^\prime+\frac{2A}{r})\partial_r[\delta F]
+(\frac{2A^\prime}{r}+\frac{2A}{r^2}+m^2)\delta F\right)
\nonumber \\
&=&(\frac{F}{r}-\frac{F^\prime}{2})A\partial_r\left[\frac{\delta
E_1}{\sqrt{A}}-\sqrt{A}\delta E_0\right]
-m^2F\frac{\delta E_1}{\sqrt{A}}\nonumber \\
&&\quad+\frac{1}{2}\left(\frac{1}{A}\partial_t^2[\delta F]
-A\partial_r^2[\delta F]
-(A^\prime+\frac{2A}{r})\partial_r[\delta F]
+(\frac{2A^\prime}{r}+\frac{2A}{r^2}+m^2)\delta F\right)=0,
\end{eqnarray}
where use has been made of (\ref{equation:Wyman F}).
One derives a simple field equation by inserting
(\ref{equation:R23 one}) in
(\ref{equation:Wyman TrRsimp}) and dropping the resulting terms
that are of
second order in the background skew field $F$:
\begin{equation}\label{equation:Wyman deltaE}
\partial_r[r\sqrt{A}\delta E]=0
\rightarrow \delta E=\frac{\delta M_s}{r\sqrt{A}},
\end{equation}
where the constant of integration has been identified with the
GR-like
perturbation of the Schwarzschild mass parameter.

Now (\ref{equation:Wyman R00plusR11}) can be used to replace
$\delta E_0$ with
$\delta E_1$ at this order, and (\ref{equation:Wyman deltaE1}) to
replace
$\delta E_1$ with $\delta E$ to find:
\begin{eqnarray}\label{equation:Wyman R23}
{}^1\! {\cal R}_{[23]}&=&\frac{1}{2}\left\{
\frac{1}{A}\partial_t^2[\delta F]
-A\partial_r^2[\delta F]
-(A^\prime+\frac{2A}{r})\partial_r[\delta F]
+(\frac{2A^\prime}{r}+\frac{2A}{r^2}+m^2)\delta
F\right\}\nonumber  \\
&&\quad-A(F^\prime-\frac{2F}{r})\partial_r[\frac{\delta E}{\sqrt{A}}]
-m^2F\frac{\delta E}{\sqrt{A}}=0,
\end{eqnarray}
and using (\ref{equation:Wyman deltaE}) in this yields the wave
equation for
$\delta F$:
\begin{eqnarray}\label{equation:Wyman deltaF}
\frac{1}{A}\partial_t^2[\delta F]
-A\partial_r^2[\delta F]
&-&(A^\prime+\frac{2A}{r})\partial_r[\delta F]
+(m^2+\frac{2A}{r^2}+\frac{2A^\prime}{r})\delta F\nonumber  \\
&=&\frac{\delta
M_s}{rA}[m^2F+4(\frac{F}{r}-\frac{F^\prime}{2})(\frac{A}{r}
+A^\prime)].
\end{eqnarray}
Note that this is a static source and so will not in itself
induce any wave
solutions, but as before the effects of a matter source will show
up
asymptotically.
The static part of the solution may be derived using the methods
in Section
\ref{section:Schw pert}:
\begin{equation}\label{equation:Wyman static dF}
\delta F=-2F_0\frac{\delta
M_s}{r}e^{-2mr^*}\ln(\frac{r^*}{2M_s}),
\end{equation}
and is consistent with the static solution (\ref{equation:Wyman
asymptF})
derived about a Schwarzschild background with mass parameter
$M_s+\delta M_s$.
Time dependent solutions are identical to those found from
(\ref{equation:Wyman
F}), and will induce time dependence in the symmetric sector
through
(\ref{equation:Wyman deltaE1}).
Since $\delta E_1$ is related to $\delta F$ locally, it is not an
independent
degree of freedom, and since the skew field is short-ranged, it
will not radiate
energy at infinity.

Using (\ref{equation:Wyman deltaF}) and (\ref{equation:Wyman F}),
one reduces
(\ref{equation:Wyman R00plusR11}) to an algebraic relation for
$\delta E_0$:
\begin{eqnarray}\label{equation:Wyman deltaE0}
&&\frac{2A}{r}\partial_r[\sqrt{A}\delta E_0+\frac{\delta
E}{\sqrt{A}}]
+AF\partial_r^2[\delta F]
+(\frac{3A^\prime}{2}F+\frac{2A}{r}F
+\frac{A}{2}F^\prime)\partial_r[\delta F]\nonumber \\
&&\quad +[(A^\prime+\frac{A}{2r})F^\prime-
(\frac{(A^\prime)^2}{2A}+\frac{7A^\prime}{2r}+\frac{3A}{r^2}
+\frac{3}{4}m^2)F]\delta F=0.
\end{eqnarray}
The solution for $\delta E_0$ can be written as:
\begin{equation}\label{equation:Wyamn deltaE0soln}
\sqrt{A}\delta E_0=\sqrt{A}\delta\tilde{E}-\frac{\delta
E}{\sqrt{A}}+B(t)
\end{equation}
where $B(t)$ is an arbitrary function of time (removable in the
usual way using
the remaining gauge transformation), the second term corresponds
to the static
Schwarzschild perturbation from Section \ref{section:Schw pert},
and
$\delta\tilde{E}$ solves the remainder of (\ref{equation:Wyman
deltaE0}).
Note that although not independent degrees of freedom, neither
$\delta E_0$ nor
$\delta E_1$ is static.
This is in fact what one would expect when considering the effect
of a
spherically symmetric matter field to which Birkhoff's theorem
does not apply,
on the GR background.
The presence of the non-static field will induce time dependence
in the
gravitational fields, without exciting any independent modes.
This is expected to continue to be the case in NGT: the general
spherically
symmetric system should only have degrees of freedom in the skew
sector.

\section{Conclusions}

The asymptotic behavior of the antisymmetric sector for the case
of a static
Wyman-type metric has been determined, and the corrections to the
symmetric
sector shown to be negligible provided one considers regions of
spacetime far
enough away from the gravitational source.
It has also been determined that if one keeps the antisymmetric
$t-r$ component,
then one cannot have asymptotic behavior that is dominated by the
Schwarzschild
metric, and so it must be discarded.
This analysis was facilitated by the introduction of a vierbein
basis, although
the formalism has been given for a general basis for
completeness.

By considering a spherically symmetric perturbation of the
Schwarzschild metric,
it has been shown that NGT does not have a rigorous Birkhoff
theorem as the
antisymmetric sector will not remain static in general.
(This has also been noted previously in a Unified Field Theory
based on Lyra
geometries \cite{Dunn:1971}.)
Perturbing an approximate Wyman background in a vierbein basis
has shown that
the symmetric sector is also not static in general, although no
additional modes
become excited.
This is important phenomenologically since one cannot consider
the static
solutions (Schwarzschild and Wyman) as the only spherically
symmetric exterior
solutions to the field equations, and one must therefore match an
interior
solution to a non-static exterior in general.

Perturbations of GR backgrounds have been shown to have good
asymptotic behavior
in general, since the ghost modes do not become excited and the
remaining
degrees of freedom are short ranged by construction.
However this is not good enough since one expects that the
physically
interesting solutions to mNGT will not be the purely GR
solutions, and one would
therefore like to examine the behavior of perturbations on
generic,
asymptotically-flat, mNGT backgrounds.
A covariant perturbative scheme, although possible in principle,
would seem to
be too complex to be of any practical value.
Instead one may treat each case separately and consider the
behavior of (perhaps
several) modes about a particular background, as was done here
for the
spherically symmetric perturbation about a Wyman background.

However this also may not be adequate to fully understand the
dynamics of the
skew sector in mNGT.
The lack of additional gauge invariance in the skew sector may
mean that there
are more modes in the rigorous theory that will be seen in any
sort of weak
field, perturbative analysis.
To determine whether or not this is the case will require a
canonical analysis
of the full theory.
Partial information may be obtained by considering the full set
of fields in a
spherically symmetric system, and looking for global information
about the
behavior of the skew modes given a general coupling to external
sources.
This is not likely to be a tractable problem in a coordinate
basis, and even in
a Lorentz frame the field equations are not expected to be
particularly
enlightening, due to their complication alone.
However the canonical analysis of this system will show which
fields propagate
in the general case, and allow one to get at the dynamics of the
approach to an
asymptotically well behaved spacetime.

\acknowledgments

The author would like to thank the Natural Sciences and
Engineering Research
Council of Canada and the University of Toronto for funding
during part of this
work, the hospitality of the Department of Physics, Cave Hill
Campus, University
of West Indies, Barbados, and the Inter America Development Bank
for supporting
the stay in Barbados.
Thanks also go to L. Demopoulos for suggestions, and J. W.
Moffat, J.
L\'{e}gar\'{e}, P. Savaria, and N. Cornish for discussions
related to this work.

\appendix

\section{Wyman Sector Field Equations}
\label{appendix:Wyman}

Here the details of the calculation of the Wyman field equations
are given,
following the steps outlined in Section \ref{section:GenBasis
intro}.
Beginning with $A^A_{\;B}=\eta^{AC}g_{[CB]}$ from (\ref{defns}),
one finds the
remaining components:
\begin{equation}
A^2_{\;3}=-A^3_{\;2}=F.
\end{equation}
The inverse of the symmetric part of the metric is the Minkowski
metric: $\eta$,
whereas the full inverse metric is:
\begin{equation}
| g^{AB} |=\left[
\begin{array}{cccc}
1&0&0&0\\
0&-1&0&0\\
0&0&-\frac{1}{1+F^2}&-\frac{F}{1+F^2}\\
0&0&\frac{F}{1+F^2}&-\frac{1}{1+F^2}\end{array}\right],
\end{equation}
and is necessary in order to compute the mass tensor
(\ref{equation:mass
tensor}).
The structure constants, and hence the skew components of the
connection
$\Gamma$, are then easily found to be:
\begin{eqnarray}
\Gamma^0_{[10]}&=&\frac{1}{2}{C_{10}}^0=\frac{1}{2}E_1\partial_r[
\ln(E_0)],
\nonumber  \\
\Gamma^2_{[12]}&=&\Gamma^3_{[13]}
=\frac{1}{2}{C_{12}}^2=\frac{1}{2}{C_{13}}^3
=-\frac{1}{2r}E_1, \nonumber \\
\Gamma^3_{[23]}&=&\frac{1}{2}{C_{23}}^3=-\frac{1}{2r}\cot(\theta)
{}.
\end{eqnarray}
Then $\Delta$ from (\ref{delta}) may be determined:
\begin{equation}
-\Delta_{1[23]}=\Delta_{3[12]}=\Delta_{2[31]}=E_1\partial_r[F],
\end{equation}
and $\Omega$ from (\ref{omega}):
\begin{eqnarray}
\Omega_{1[23]}&=&E_1(\frac{F}{r}-\frac{1}{2}\partial_r[F]),
\nonumber \\
 \Omega_{3[12]}&=&\Omega_{2[31]}=\case{1}{2}E_1\partial_r[F].
\end{eqnarray}

Now (\ref{compantisymm}) is solved for $\Lambda$ and
(\ref{symmconn}) to
calculate the connection components ($\Gamma$):
\begin{eqnarray}
\Lambda^3_{12}&=&\Lambda^2_{31}=-\frac{1}{2}\frac{E_1F^\prime}{1+
F^2}, \nonumber \\
\Lambda^1_{23}&=&-\frac{E_1F}{r}+\frac{1}{2}E_1F^\prime-\frac{1}{
2}E_1F\partial_r
[\ln(1+F^2)],
\end{eqnarray}
and:
\begin{eqnarray}
\Gamma^0_{01}&=&\Gamma^1_{00}=-E_1\partial_r[\ln(E_0)], \nonumber
\\
\Gamma^1_{22}&=&\Gamma^1_{33}=-\frac{E_1}{r}
-\frac{1}{2}E_1\partial_r[\ln(1+F^2)]
, \nonumber \\
\Gamma^2_{21}&=&\Gamma^3_{31}
=\frac{E_1}{r}+\frac{1}{4}E_1\partial_r[\ln(1+F^2)] ,\nonumber \\
\Gamma^2_{12}&=&\Gamma^3_{13}
=\case{1}{4}E_1\partial_r[\ln(1+F^2)] , \nonumber \\
\Gamma^3_{32}&=&-\Gamma^2_{33}=\frac{1}{r}\cot(\theta).
\end{eqnarray}

The field equations that remain are:
\begin{eqnarray}\label{equation:Wyman field equations}
{\cal R}_{00}&=&E_1\partial_r[\Gamma^1_{00}]
+\Gamma^1_{00}(\Gamma^0_{01}+2\Gamma^2_{21})
+\frac{m^2}{4}\frac{F^2}{1+F^2}=0,\nonumber  \\
{\cal
R}_{11}&=&-E_1\partial_r[\Gamma^0_{01}]-2E_1\partial_r[\Gamma^2_{
21}]
-(\Gamma^0_{01})^2-2(\Gamma^2_{21})^2-2(\Lambda^3_{12})^2
-\frac{m^2}{4}\frac{F^2}{1+F^2}=0,\nonumber  \\
{\cal R}_{22}&=&{\cal
R}_{33}=E_1\partial_r[\Gamma^1_{22}]+\frac{1}{r^2}
+\Gamma^1_{22}(\Gamma^0_{01}+4\Gamma^2_{[21]})
-2\Lambda^3_{12}\Lambda^1_{23}+\frac{m^2}{4}\frac{F^2}{1+F^2}=0,
\nonumber  \\
{\cal R}_{[23]}&=&E_1\partial_r[\Lambda^1_{23}]
+(\Gamma^0_{01}-4\Gamma^2_{[12]})\Lambda^1_{23}
+2\Gamma^1_{22}\Lambda^3_{12}
-\frac{m^2}{4}(F+\frac{F}{1+F^2})=0.
\end{eqnarray}
It can be shown that in the absence of any skew sector
altogether, the
Schwarzschild field equations are obtained, and in the absence of
the mass term
one has the Wyman Field equations \cite{Wyman:1950}.

\section{Perturbation Equations}
\label{appendix:Wyman perturbation}

The background field will be that given in Appendix
\ref{appendix:Wyman},
approximated by considering only first order contributions from
the skew sector.
The spherically symmetric perturbations about this background are
given here in
detail, keeping only first order in the background $F$, and
setting $\alpha=3/4$
throughout.
One may begin by calculating first order corrections to metric
quantities, first
the density:
\begin{equation}
\delta\sqrt{-g}=\sqrt{-g}F\delta F,
\end{equation}
and the inverse of the full metric metric is:
\begin{equation}
| \delta g^{AB} |=\left[
\begin{array}{cccc}
0&\delta W&0&0\\
-\delta W&0&0&0\\
0&0&2F\delta F&-\delta F\\
0&0&\delta F&2F\delta F\end{array}\right].
\end{equation}
The tensor $\delta A$ has remaining components:
\begin{equation}
\delta A^0_{\;1}=\delta A^1_{\;0}=\delta W,\quad
\delta A^2_{\;3}=-\delta A^3_{\;2}=-\delta F,
\end{equation}
and the perturbation of the antisymmetric connection
coefficients, derived from
(\ref{Stcnst}):
\begin{eqnarray}
\delta\Gamma^0_{[10]}&=&\frac{1}{2}\delta C_{10}^{\;\;\;\;0}
=\frac{1}{2}(\delta
E_1\partial_r[\ln(E_0)]+\frac{E_1}{E_0}\partial_r[\delta
E_0]
-\frac{E_1}{E_0}\partial_r[\ln(E_0)]\delta E_0), \nonumber \\
\delta\Gamma^1_{[01]}&=&\frac{1}{2}\delta C_{01}^{\;\;\;\;1}
=\frac{1}{2}\frac{E_0}{E_1}\partial_t[\delta E_1],\nonumber  \\
\delta\Gamma^2_{[12]}&=&\delta\Gamma^3_{[13]}
=\frac{1}{2}\delta C_{12}^{\;\;\;\;2}=\frac{1}{2}\delta
C_{13}^{\;\;\;\;3}
=-\frac{1}{2}\frac{\delta E_1}{r}.
\end{eqnarray}

The last equation of (\ref{mNGTF}) is now solved to determine
$\delta W$ in
terms of
metric functions to find:
\begin{eqnarray}
\delta W_0&=&\frac{4}{3}E_1(\partial_r[\delta W]
+\frac{2\delta W}{r}) ,\nonumber \\
\delta W_1&=&\case{4}{3}E_0\partial_t[\delta W] ,\nonumber \\
\delta W_2&=&\delta W_3=0.
\end{eqnarray}

It is then fairly straightforward to calculate the remaining
$\delta\Delta^0$:
\begin{eqnarray}
\delta\Delta^0_{0[01]}&=&-\case{1}{4}\delta W_1,\nonumber \\
\delta\Delta^0_{1[01]}&=&-\case{1}{4}\delta W_0,\nonumber \\
\delta\Delta^0_{2[a2]}&=&\delta\Delta^0_{3[a3]}
=-\case{1}{4}\delta W_a, \nonumber \\
\delta\Delta^0_{2(3a)}&=&-\delta\Delta^0_{3(2a)}
=\case{1}{4}F\delta W_a,
\end{eqnarray}
where $a\in\{0,1\}$ here and in the following.
The remaining $\Delta$s are:
\begin{eqnarray}
\delta\Delta_{2(a3)}&=&-\delta\Delta_{3(a2)}
=-\case{1}{2}F\delta W_a ,\nonumber \\
\delta\Delta_{0[10]}&=&
2E_0\partial_t[\delta W]-\case{1}{2}\delta W_1,\nonumber  \\
\delta\Delta_{1[10]}&=&
2E_1\partial_r[\delta W]-\case{1}{2}\delta W_0 ,\nonumber \\
\delta\Delta_{2[2a]}&=&\delta\Delta_{3[3a]}=
-\case{1}{2}\delta W_a ,\nonumber \\
\delta\Delta_{0[23]}&=&-E_0\partial_t[\delta F], \nonumber \\
\delta\Delta_{1[23]}&=&
-E_1\partial_r[\delta F]-\delta E_1\partial_r[F] ,\nonumber \\
\delta\Delta_{3[02]}&=&-\delta\Delta_{2[03]}=
E_0\partial_t[\delta F] ,\nonumber \\
\delta\Delta_{3[12]}&=&-\delta\Delta_{2[13]}=
E_1\partial_r[\delta F]+\delta E_1\partial_r[F].
\end{eqnarray}
Then $\delta\Omega$ can be calculated:
\begin{eqnarray}
\delta\Omega_{0[10]}&=&
E_0\partial_t[\delta W]-\case{1}{4}\delta W_1 ,\nonumber \\
\delta\Omega_{1[10]}&=&
E_1\partial_r[\delta W]-\case{1}{4}\delta W_0 ,\nonumber \\
\delta\Omega_{2[02]}&=&\delta\Omega_{3[03]}=
\frac{1}{4}\delta W_0
-\frac{E_1\delta W}{r} ,\nonumber \\
\delta\Omega_{2[12]}&=&\delta\Omega_{3[13]}=
\case{1}{4}\delta W_1 ,\nonumber \\
\delta\Omega_{0[23]}&=&
-\case{1}{2}E_0\partial_t[\delta F] ,\nonumber \\
\delta\Omega_{1[23]}&=&
-\frac{1}{2}E_1\partial_r[\delta F]+\frac{F\delta
E_1}{r}+\frac{E_1\delta F}{r}
-\frac{1}{2}\delta E_1\partial_r[F],\nonumber \\
\delta\Omega_{3[02]}&=&-\delta\Omega_{2[03]}=
\case{1}{2}E_0\partial_t[\delta F] ,\nonumber \\
\delta\Omega_{3[12]}&=&-\delta\Omega_{2[13]}=
\case{1}{2}E_1\partial_r[\delta F]+\case{1}{2}\delta
E_1\partial_r[F].
\end{eqnarray}

One is now in a position to invert the compatibility conditions
(\ref{compantisymm}) to solve for the perturbations to the
connection
coefficients.
First the corrections to $\Lambda$:
\begin{eqnarray}
\delta\Lambda^0_{01}&=&-\case{2}{3}E_0\partial_t[\delta W],
\nonumber \\
\delta\Lambda^1_{01}&=&-\frac{2}{3}E_1(\frac{\delta
W}{r}-\partial_r[\delta W]) ,
\nonumber\\
\delta\Lambda^2_{20}&=&\delta\Lambda^3_{30}=
-\frac{1}{3}E_1(\frac{\delta W}{r}-\partial_r[\delta W])
,\nonumber \\
\delta\Lambda^2_{21}&=&\delta\Lambda^3_{31}=
\case{1}{3}E_0\partial_t[\delta W] , \nonumber\\
\delta\Lambda^0_{23}&=&-\case{1}{2}E_0\partial_t[\delta
F],\nonumber \\
\delta\Lambda^1_{23}&=&-\frac{E_1\delta F}{r}-\frac{\delta
E_1F}{r}
+\frac{1}{2}\frac{\delta E_1F^\prime}{r}
+\frac{1}{2}\frac{E_1\partial_r[\delta F]}{r} ,\nonumber \\
\delta\Lambda^2_{03}&=&\delta\Lambda^3_{20}=
\case{1}{2}E_0\partial_t[\delta F] ,\nonumber \\
\delta\Lambda^2_{13}&=&\delta\Lambda^3_{21}=
\case{1}{2}E_1\partial_r[\delta F]
+\case{1}{2}F^\prime\delta E_1,
\end{eqnarray}
and then to $\Gamma$:
\begin{eqnarray}
\delta\Gamma^0_{01}&=&\delta\Gamma^1_{00}=
-\frac{E_1}{E_0}\partial_r[\delta E_0]
-\delta E_1\partial_r[\ln(E_0)]
+\frac{E_1}{E_0}\delta E_0\partial_r[\ln(E_0)] ,\nonumber \\
\delta\Gamma^0_{11}&=&\delta\Gamma^1_{10}=
-\frac{E_0}{E_1}\partial_t[\delta E_1] ,\nonumber \\
\delta\Gamma^2_{03}&=&\delta\Gamma^2_{30}=
-\delta\Gamma^3_{02}=-\delta\Gamma^3_{20}=
\case{1}{2}E_1 F^\prime\delta W ,\nonumber \\
\delta\Gamma^0_{22}&=&\delta\Gamma^0_{33}=
2\delta\Gamma^2_{02}=2\delta\Gamma^2_{20}=
2\delta\Gamma^3_{03}=2\delta\Gamma^3_{30}=
E_0 F\partial_t[\delta F] , \nonumber\\
\delta\Gamma^1_{22}&=&\delta\Gamma^1_{33}=
-\frac{\delta E_1}{r}
-E_1 F\partial_r[\delta F]
-E_1F^\prime\delta F ,\nonumber \\
\delta\Gamma^2_{12}&=&\delta\Gamma^3_{13}=
\case{1}{2}E_1 F\partial_r[\delta F]
+\case{1}{2}E_1F^\prime\delta F ,\nonumber \\
\delta\Gamma^2_{21}&=&\delta\Gamma^3_{31}=
\frac{\delta E_1}{r}
+\frac{1}{2}E_1 F\partial_r[\delta F]
+\frac{1}{2}E_1F^\prime\delta F.
\end{eqnarray}
The remaining field equations will be: ${}^1\! {\cal
R}_{00},{}^1\! {\cal R}_{(01)}, {}^1\! {\cal
R}_{[01]}, {}^1\! {\cal R}_{11},{}^1\! {\cal
R}_{22}={}^1\! {\cal R}_{33}$ and ${}^1\! {\cal
R}_{[23]}$.
The relevant combinations will be quoted in Section
\ref{section:Wyman pert}.

\bibliographystyle{prabib}

\begin{references}

\bibitem{Moffat:1994}
J.~W. Moffat.
\newblock Nonsymmetric gravitational theory.
\newblock {\em J. Math. Phys.}, 36, 3722--3732, 1995.

\bibitem{Moffat:1995b}
J.~W. Moffat.
\newblock A new nonsymmetric gravitational theory.
\newblock {\em Phys. Lett. B}, 355, 447--452, 1995.

\bibitem{Regge+Wheeler:1957}
T.~Regge and J.~A. Wheeler.
\newblock Stability of the {S}chwarzschild singularity.
\newblock {\em Phys. Rev. D}, 108, 1063--1069, 1957.

\bibitem{Vishveshwara:1970}
C.~V. Vishveshwara.
\newblock Stability of the {S}chwarzschild metric.
\newblock {\em Phys. Rev. D}, 1, 2870--2879, 1970.

\bibitem{Kalb+Ramond:1974}
M.~Kalb and P.~Ramond.
\newblock Classical direct interstring action.
\newblock {\em Phys. Rev. D}, 9, 2273--2284, 1974.

\bibitem{Isenberg+Nestor:1977}
J.~A. Isenberg and J.~M. Nestor.
\newblock The effect of a gravitational interaction on classical
fields: A
  {H}amiltonian-{D}irac analysis.
\newblock {\em Ann. Phys.}, 107, 56--81, 1977.

\bibitem{Moffat:1979}
J.~W. Moffat.
\newblock New theory of gravitation.
\newblock {\em Phys. Rev. D}, 19, 3554--3558, 1979.

\bibitem{Einstein:1945}
A.~Einstein.
\newblock A generalization of the relativistic theory of
gravitation.
\newblock {\em Ann. Math.}, 46, 578--584, 1945.

\bibitem{Einstein+Straus:1946}
A.~Einstein and E.~G. Straus.
\newblock A generalization of the relativistic theory of
gravitation, ii.
\newblock {\em Ann. Math.}, 47, 731--741, 1946.

\bibitem{Moffat:1990}
J.~W. Moffat.
\newblock Review of the nonsymmetric gravitational theory.
\newblock In R.~Mann and P.~Wesson, editors, {\em Gravitation: A
Banff Summer
  Institute}, New Jersey, 1991. World Scientific.

\bibitem{Einstein:1956}
Albert Einstein.
\newblock {\em The Meaning of Relativity}.
\newblock Princeton University Press, Princeton, New Jersey, 5
edition, 1974.

\bibitem{Kunstatter+Yates:1981}
G.~Kunstatter and R.~Yates.
\newblock The geometrical structure of a complexified theory of
gravitation.
\newblock {\em J. Phys. A}, 14, 847--854, 1981.

\bibitem{Mann:1989}
R.~B. Mann.
\newblock New ghost-free extensions of general relativity.
\newblock {\em Class. Quantum Grav.}, 6, 41--57, 1989.

\bibitem{Kunstatter+Moffat+Malzan:1983}
G.~Kunstatter, J.~W. Moffat, and J.~Malzan.
\newblock Geometrical interpretation of a generalized theory of
gravitation.
\newblock {\em J. Math. Phys.}, 24, 886--889, 1983.

\bibitem{Kelly:1991}
P.~F. Kelly.
\newblock Expansions of non-symmetric gravitational theories
about a {GR}
  background.
\newblock {\em Class. Quantum Grav.}, 8, 1217--1229, 1991.

\bibitem{Wald:1984}
Robert~M. Wald.
\newblock {\em General Relativity}.
\newblock The University of Chicago Press, Chicago, 1984.

\bibitem{Damour+Deser+McCarthy:1992}
T.~Damour, S.~Deser, and J.~McCarthy.
\newblock Theoretical problems in nonsymmetric gravitational
theory.
\newblock {\em Phys. Rev. D}, 45, R3289--R3291, 1992.

\bibitem{Damour+Deser+McCarthy:1993}
T.~Damour, S.~Deser, and J.~McCarthy.
\newblock Nonsymmetric gravity theories: Inconsistencies and a
cure.
\newblock {\em Phys. Rev. D}, 47, 1541--1556, 1993.

\bibitem{Kelly:1992}
P.~F. Kelly.
\newblock Expansions of non-symmetric gravitational theories
about a {GR}
  background.
\newblock {\em Class. Quantum Grav.}, 9, 1423, 1992.
\newblock Erratum.

\bibitem{Fischer+Marsden:1979}
A.~E. Fischer and J.~E. Marsden.
\newblock The initial value problem and the dynamical formulation
of general
  relativity.
\newblock In S.~W. Hawking and W.~Israel, editors, {\em General
Relativity: An
  Einstein Centenary Survey}, New York, 1979. Cambridge
University Press.

\bibitem{Nakanishi:1967}
N.~Nakanishi.
\newblock Quantum electrodynamics in the general covariant gauge.
\newblock {\em Prog. theor. Phys.}, 38, 881--891, 1967.

\bibitem{Goto+Obara:1967}
T.~Goto and T.~Obara.
\newblock The canonical quantization of the free electromagnetic
field in the
  landau gauge.
\newblock {\em Prog. Theor. Phys.}, 38, 871--880, 1967.

\bibitem{vanNieuw:1973}
P.~vanNieuwenhuizen.
\newblock On ghost-free tensor {L}agrangians and linearized
gravitation.
\newblock {\em Nucl. Phys. B}, 60, 478--492, 1973.

\bibitem{Kunstatter+Leivo+Savaria:1984}
G.~Kunstatter, H.~P. Leivo, and P.~Savaria.
\newblock Dirac constraint analysis of a linearized theory of
gravitation.
\newblock {\em Class. Quantum Grav.}, 1, 7--13, 1984.

\bibitem{Mann+Moffat:1982}
R.~B. Mann and J.~W. Moffat.
\newblock Ghost properties of generalized theories of
gravitation.
\newblock {\em Phys. Rev. D}, 26, 1858--1861, 1982.

\bibitem{Moffat:1980}
J.~W. Moffat.
\newblock A solution of the {C}auchy initial value problem in the
nonsymmetric
  theory of gravitation.
\newblock {\em J. Math. Phys.}, 21, 1798--1801, 1980.

\bibitem{McDow+Moffat:1982}
J.~C. McDow and J.~W. Moffat.
\newblock Consistency of the {C}auchy initial value problem in a
nonsymmetric
  theory of gravitation.
\newblock {\em J. Math. Phys.}, 23, 634--636, 1982.

\bibitem{Moffat:1981}
J.~W. Moffat.
\newblock Gauge invariance and string interactions in a
generalized theory of
  gravitation.
\newblock {\em Phys. Rev. D}, 23, 2870--2874, 1981.

\bibitem{Mann:1986}
R.~B. Mann.
\newblock Gravity, ghosts, and strings.
\newblock {\em Can. J. Phys.}, 64, 589--594, 1985.

\bibitem{Cornish+Moffat:1994}
N.~J. Cornish and J.~W. Moffat.
\newblock A non-singular theory of gravity?
\newblock {\em Phys. Lett. B}, 336, 337--342, 1994.

\bibitem{Bogoliubov+Shirkov:1959}
N.~N. Bogoliubov and D.~V. Shirkov.
\newblock {\em Introduction to the Theory of Quantized Fields}.
\newblock Interscience Publishers inc., New York, 1959.

\bibitem{Itzykson+Zuber:1980}
C.~Itzykson and J.-B. Zuber.
\newblock {\em Quantum Field Theory}.
\newblock McGraw-Hill Book Company, Toronto, 1980.

\bibitem{Reed+Simon:1979}
M.~Reed and B.~Simon.
\newblock {\em Scattering Theory}, volume III of {\em Methods of
Modern
  Mathematical Physics}.
\newblock Academic Press, Inc., New York, 1979.

\bibitem{Tonnelat:1982}
M.~A. Tonnelat.
\newblock {\em {E}instein's Theory of Unified Fields}.
\newblock Gordon and Breach, Science Publishers, New York, 1982.

\bibitem{Vlachynsky:1988}
Eugen~Josef Vlachynsky.
\newblock {\em Analytic Solutions in {M}offat's Nonsymmetric
Generalized Theory
  of Gravitation}.
\newblock PhD thesis, Department of Applied Mathematics,
University of Sydney,
  1988.

\bibitem{Hlavaty:1958}
V.~Hlavat\'{y}.
\newblock {\em Geometry of {E}instein's Unified Field Theory}.
\newblock P. Noordhoff Ltd., Groningen, Holland, 1958.

\bibitem{Choquet-Bruhat+:1989}
Y.~Choquet-Bruhat, C.~DeWitt-Morette, and M.~Dillard-Bleik.
\newblock {\em Analysis, Manifolds and Physics}, volume~1.
\newblock North Holland, New York, 1989.

\bibitem{Nakahara:1990}
M.~Nakahara.
\newblock {\em Geometry, Topology and Physics}.
\newblock Adam Hilger, New York, 1990.

\bibitem{Moffat:1987}
J.~W. Moffat.
\newblock Test-partical motion in the nonsymmetric gravitation
theory.
\newblock {\em Phys. Rev. D}, 35, 3733--3747, 1987.

\bibitem{Legare+Moffat:1995}
J.~L\'{e}gar\'{e} and J.~W. Moffat.
\newblock Field equations and conservation laws in the
nonsymmetric
  gravitational theory.
\newblock {\em Gen. Rel. Grav.}, 27, 761--775, 1995.

\bibitem{Wyman:1950}
M.~Wyman.
\newblock Unified field theory.
\newblock {\em Can. J. Math.}, 2, 427--439, 1950.

\bibitem{Cornish:1994}
N.~J. Cornish.
\newblock The nonsingular {S}chwarzschild-like solution to
massive nonsymmetric
  gravity.
\newblock {\em UTPT-94-37}, 1994.

\bibitem{Dunn:1971}
Kenneth~A. Dunn.
\newblock {\em {B}irkhoff's Theorem and Equations of Motion in a
Unified Field
  Theory}.
\newblock PhD thesis, Department of Mathematics, University of
Toronto, 1971.

\end{references}


\end{document}